\documentclass[10pt,journal,compsoc]{IEEEtran}
\usepackage{multirow, rotating, amsmath, wasysym}
\usepackage{tabularx}
\usepackage{wrapfig}
\usepackage{booktabs}
\usepackage{epstopdf}
\usepackage[tight]{subfigure}
\usepackage{algpseudocode}
\usepackage{adjustbox}
\usepackage{graphicx}
\usepackage{caption}
\usepackage{algorithm}
\usepackage{soul,xcolor}
\usepackage{todonotes}
\usepackage{balance}

\mathchardef\Gamma="0100 \mathchardef\Delta="0101
\mathchardef\Theta="0102 \mathchardef\Lambda="0103
\mathchardef\Xi="0104 \mathchardef\Pi="0105
\mathchardef\Sigma="0106 \mathchardef\Upsilon="0107
\mathchardef\Phi="0108 \mathchardef\Psi="0109
\mathchardef\Omega="010A

\newcommand{\outline}[1]{}%{\textbf{#1}}

\usepackage{xspace}
\usepackage{url}
\usepackage{graphicx}
\usepackage{latexsym}
\usepackage{amssymb}
\usepackage{amsfonts}
\usepackage{psfrag}
\usepackage{wrapfig}
\usepackage{comment}
\usepackage{alltt}
\usepackage{color}

\newcommand{\ie}{\emph{i.e.}\xspace}
\newcommand{\eg}{\emph{e.g.}\xspace}

\newcommand{\etal}{\frenchspacing{}\emph{et al{.}}\xspace}
\newcommand{\presec}{\vspace{-0.03in}}
\newcommand{\postsec}{\vspace{-0.03in}}
\newcommand{\presub}{\vspace{-0.03in}}
\newcommand{\postsub}{\vspace{-0.03in}}

\begin{document}
\title{On Goodness of WiFi based Monitoring of Vital Signs in the Wild}

\author{Kamran~Ali, Mohammed Alloulah, Fahim Kawsar, Alex~X.~Liu
    \IEEEcompsocitemizethanks{\IEEEcompsocthanksitem K. Ali and A. X. Liu are with the Department Computer Science and Engineering, Michigan State University, Lansing, MI, USA, 48823.\protect\\
        E-mail: {alikamr3, alexliu}@cse.msu.edu 
        \IEEEcompsocthanksitem M. Alloulah and F. Kawsar are with the Nokia Bell Labs, Cambridge, UK.\protect\\
        E-mail: {mohammed.alloulah, fahim.kawsar}@nokia-bell-labs.com
    }%
}

\IEEEtitleabstractindextext{
    \begin{abstract}
	WiFi channel state information (CSI) has emerged as a plausible modality for sensing different human vital signs, i.e. respiration and body motion, as a function of modulated wireless signals that travel between WiFi devices. 
	Although a remarkable proposition, most of the existing research in this space struggles to withstand robust performance beyond experimental conditions. 
	To this end, we take a careful look at the dynamics of WiFi signals under human respiration and body motions in the wild. 
	We first characterize the WiFi signal components - multipath and signal subspace -  that are modulated by human respiration and body motions. 
	We extrapolate on a set of transformations, including first-order differentiation, max-min normalization and component projections, that faithfully explains and quantifies the dynamics of respiration and body motions on WiFi signals. 
	Grounded in this characterization, we propose two methods: 1) a respiration tracking technique that models the peak dynamics observed in the time-varying signal subspaces and 2) a body-motion tracking technique built with a multi-dimensional clustering of evolving signal subspaces. 
	Finally, we reflect on the manifestation of these techniques in a practical sleep monitoring application.
	Our systematic evaluation with over 550 hours of data from 5 users covering both line-of-sight (LOS) and non-line-of-sight (NLOS) settings shows that the proposed techniques can achieve comparable performance to purpose-built pulse-Doppler radar.
\end{abstract}

}

\maketitle

\thispagestyle{empty}

\IEEEdisplaynontitleabstractindextext

\IEEEpeerreviewmaketitle

\sloppy{
 \presec
\section{Introduction}
Respiration rate and body motions are critical indicators of an individual's general state of health. 
They carry meaningful insights to assess different cardiovascular, neurological, and psychiatric functions of the human body and play an essential role in early diagnosis of various medical conditions, including sleep apnea, asthma, nausea, and several others. 
Most of the technologies that can monitor respiration and body motions simultaneously are invasive and require the subject to be connected to the measuring equipment, e.g., a respiratory inductance plethysmography belt or multiple wearable sensors.  While these instruments certainly offer medical-grade insights, they are not suitable for long-term usage due to their poor ergonomics that hinder long-term assessment.   
Several other less obtrusive methods have been proposed for tracking vital signs, for instance, Actigraphy \cite{sadeh1995role,sadeh2002role, ancoli2003role, long2014sleep, fitbit, miband, ouraring} and EEG \cite{zeo} based techniques. However, these methods still require body contact, which is something people are often not comfortable with \cite{choe2010opportunities}.

Naturally, contact-less vital signs monitoring technologies have attracted significant interest, which mainly include Audio \cite{hao2013isleep, pevernagie2010acoustics, de2009detection}, Video \cite{heinrich2015video, poh2011advancements}, Bed sensors (\eg Ballistocardiography (BCG), pressure and/or motion sensors based techniques) \cite{migliorini2010automatic, paalasmaa2012unobtrusive, kortelainen2010sleep, chen2005unconstrained, choi2009slow, withingssleeppad} and RF sensing based techniques - e.g., mmWave, Frequency Modulated Continuous Wave (FMCW) radar, Pulse-Doppler radar, RFIDs and WiFi based techniques \cite{rahman2015dopplesleep, liu2014wi, liu2015tracking, yang2016monitoring, adib2015smart, yue2018extracting, occhiuzzi2010rfid, occhiuzzi2014night}.
RF-based techniques are by far some of the least intrusive methods, both in terms of privacy and convenience of use.
Radar-based techniques can monitor breathing and other movements reasonably well, however, their operation often requires line-of-sight (LOS) which leads to deployment complexity and significant directivity issues. 
In contrast, WiFi, and in particular channel state information (CSI) signals of WiFi have emerged as an attractive modality to track respiration and body motions \cite{hillyard2018experience, liu2015tracking, liu2014wi, wang2016human, zhang2018fresnel}. 
The fundamental principle of these works is to model the variation of wireless signals modulated by the respiration and motion of a human body. 
These works have shown the remarkable ability to re-purpose WiFi signals to track vital signs; however, unfortunately, often under constrained and controlled settings with strict assumptions. 
For example, the techniques proposed in existing works have been designed based on controlled experiments often performed on the same subject, where they require the subject to lie down in between or very close to both transmitter (TX) and the receiver (RX) to ensure line-of-sight (LOS) scenarios.
Their techniques rely on trial-and-error based positioning of WiFi transceivers and signal processing methods to track vital signs, which often leads to high dependency on multiple, environment-dependent parameters that are difficult to tune in real life. 
Such techniques may be suitable for controlled short-duration lab experiments. 
However, their suitability cannot be generalized to different individuals, environments, positioning of WiFi transceivers, LOS/NLOS situations, and to natural in-home monitoring scenarios.

Building on the existing WiFi based vital signs monitoring works and recognizing their aforementioned limitations, in this work, we take a close look at the dynamics of WiFi, human respiration, and body motions in the wild. 
Extrapolating on a set of transformations including first-order differentiation, min-max normalization and component projections, first, we analyze the impact of respiration on the multipath components of WiFi signal to quantify the effect of small breathing movements on the CSI signals. 
Second, we analyze the impact of respiration on WiFi signals subspace to quantify how breathing affects the spatio-frequency subspace formed by multiple TX-RX antennas (MIMO) and Orthogonal Frequency Division Multiplexing (OFDM) subcarriers very differently compared to other bodily motions (such as slight head or limb movements). 
Collectively, these characterizations enable us to develop two robust methods for tracking respiration and body motions without any constrains. 
Moreover, these techniques eliminate user- and environment-specific calibration efforts and as such, allow us to build a system that can track vital signs using design-time training data obtained from only a few configurations and users.

We systematically evaluate our methods by first taking sleep monitoring as a case study, where we collected more than 550 hours (80 nights) of data \footnote{The dataset and deployment settings will be made public.} from 5 users at their respective apartments in real-world full-night sleep monitoring settings.
Our experiments covered both line-of-sight (LOS) and non-line-of-sight (NLOS) scenarios such that ~55\% of our dataset corresponds to NLOS deployment scenarios, and ~45\% to LOS.
Second, we develop a system named \textit{Serene} that implements our proposed methods to track vital signs during sleep.
We evaluate Serene's performance in terms of \textit{breath rate error}, number of \textit{motion false positives} that occur in a user's environment while the user is sleeping, and \textit{breath signal outage} during which Serene cannot track subject's vital signs but the ground truth device can.
Our results demonstrate that the proposed techniques were able to track respiration rate with an average error of \textless1.19 breaths per minute (BPM).
However, the breath rate error varied between 0.34 BPM to more than 5 BPM depending upon the time of night as a user's sleep posture and distance from the sleep monitor can change during sleep.
Serene experienced 20 false positive motion events on average every night, which can be attributed to activity of other house residents while a user is sleeping. 
Although the total duration of such events during any night stayed below 10 minutes on average and below 37 minutes 95\% of the time, yet we observed motion false positives of more than 60 minutes (1 hour) in total during one of the nights in our dataset.
Serene experienced an average nightly breath signal outage of 6.38 minutes.
Such outages arise due to subjects rolling over in bed to a different position and/or sleep posture that makes it difficult for Serene to pick up the subjects' chest movements for a while.
Figures \ref{fig:breathingexample} and \ref{fig:motionexample} show how Serene tracks breathing and body movement during a full night's sleep of a subject, when the WiFi sleep monitor was placed on a table close to the subject's bed and the router was placed in their TV lounge.

\begin{figure*}[htbp]
    \centering
    \captionsetup{justification=centering}
    \subfigure[Respiration tracking (full night's sleep)]{
        \includegraphics[width=0.495\textwidth]{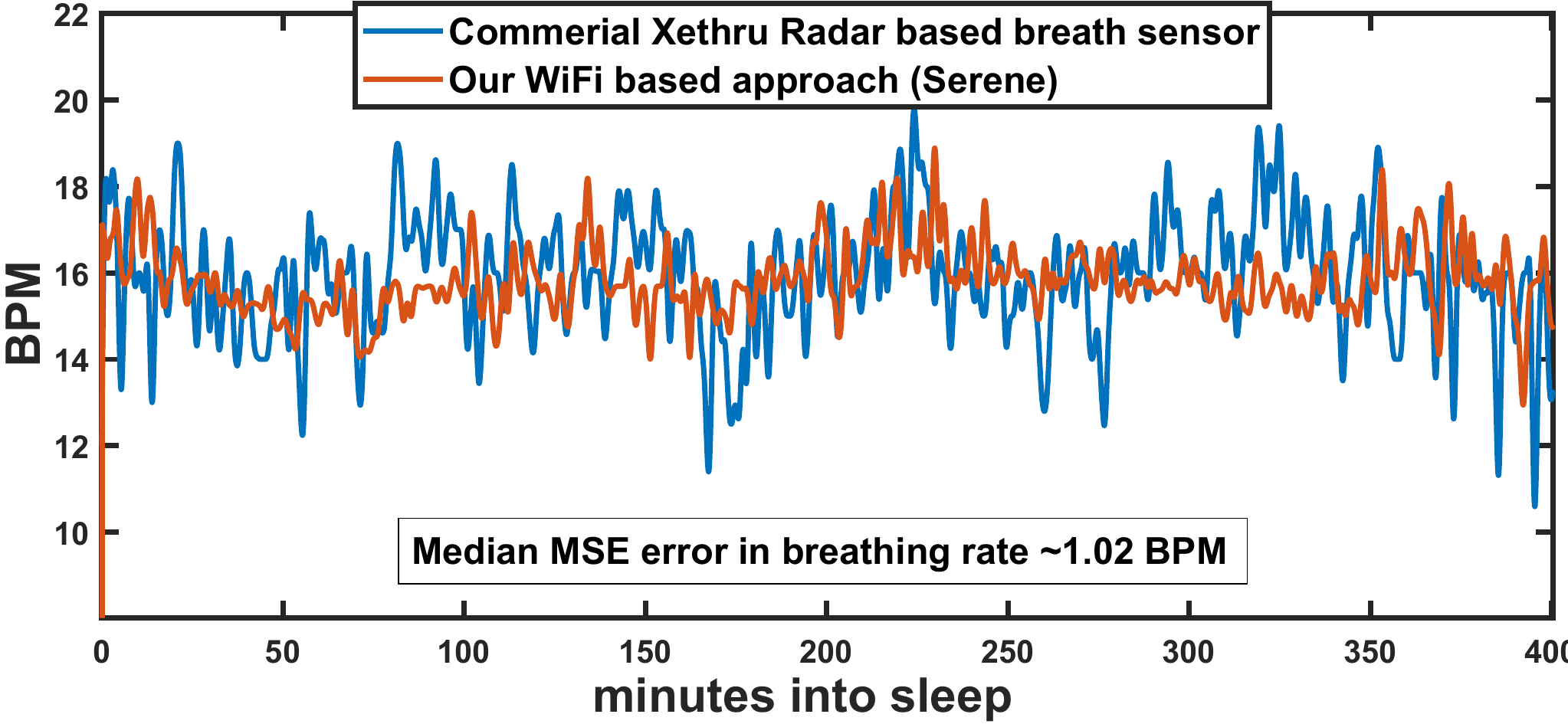}
        \label{fig:breathingexample}
    }
    \subfigure[Body movement tracking (full night's sleep)]{
        \includegraphics[width=0.415\textwidth]{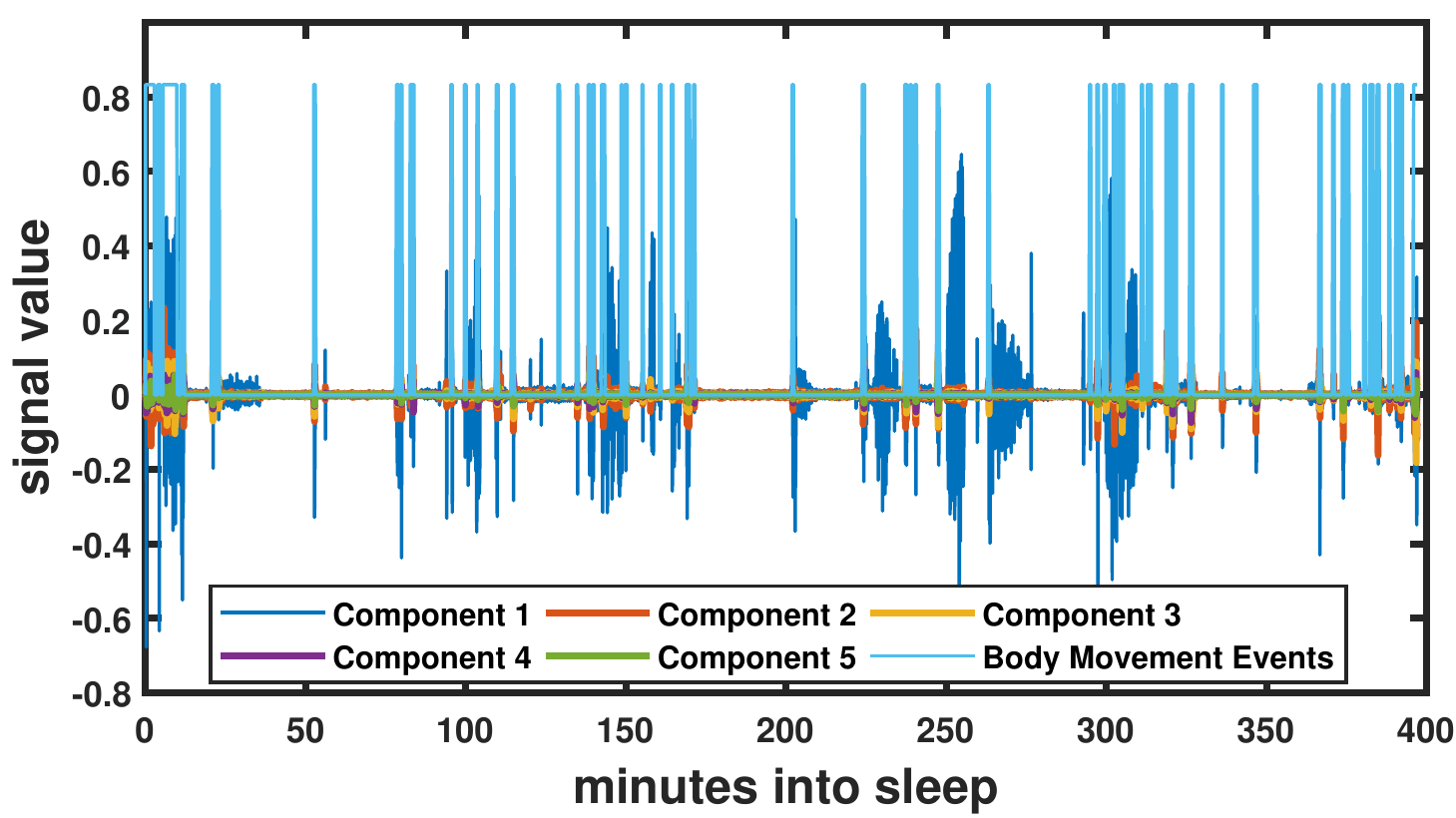}
        \label{fig:motionexample}
    }
    \vspace{-0.15in}
    \caption{Example showing our system tracking breathing and body movements throughout full night's sleep of subject. }
    \label{fig:R3_behavioralclusters}
    \vspace{-0.12in}
\end{figure*}

In what follows, we first position our work against the existing body of research. 
Then we present the characterization of the dynamics of respiration and body motion concerning modulated WiFi signals. 
The signal processing steps and implementation details are then discussed. 
We then explain our data collection and deployment configurations, followed by the evaluation reports. 
We conclude the paper by discussing a sleep efficiency monitoring application and highlight the limitations of WiFi based vital signs monitoring in the wild.
 \presec
\section{Related Work}
\postsec

\subsection{Respiration, Body Movements and Sleep}
\postsec
Previous works have shown that breathing and body movements during sleep are closely related to sleep quality in humans \cite{long2015analysis,sadeh2002role,dafna2015sleep}. 
These studies show that respiratory dynamics vary over sleep stages, which means that respiratory activity can be used to separate sleep stages \cite{long2015analysis}.
For example, Dafna \etal evaluated whole night sleep based on sleep-awake classification using audio recordings of breathing sounds \cite{dafna2015sleep}.
They captured and quantified variations in breathing features such as periodicity and consistency, and showed that these features contribute to distinguishing between sleep and wake epochs.
Our work is motivated by such studies, where our goal is to first develop a robust and generic scheme to extract breathing and limb/body activity related vital signs using CSI signals obtained from COTS WiFi devices, and then evaluate it in practical sleep experiments.

\presec
\subsection{Sleep Monitoring Technologies}
\postsec
Several sleep monitoring techniques have been proposed in the past which use different sensing modalities, such as in-ear~\cite{nguyen2017libs}, inertial sensors (Actigraphy) \cite{sun2017sleepmonitor, sadeh1995role,sadeh2002role, ancoli2003role, long2014sleep, fitbit, miband, ouraring}, EEG \cite{zeo}, Audio \cite{hao2013isleep, pevernagie2010acoustics, de2009detection}, Video \cite{heinrich2015video, poh2011advancements}, Bed sensors (\eg Ballistocardiography (BCG), pressure and/or motion sensors based techniques) \cite{migliorini2010automatic, paalasmaa2012unobtrusive, kortelainen2010sleep, chen2005unconstrained, choi2009slow, withingssleeppad} and RF sensing based techniques (\eg mmWave, Frequency Modulated Continuous Wave (FMCW) radar, Pulse-Doppler radar, RFID and WiFi based techniques) \cite{occhiuzzi2010rfid, occhiuzzi2014night, rahman2015dopplesleep, liu2014wi, liu2015tracking, yang2016monitoring, adib2015smart, yue2018extracting}.
For brevity, we will only discuss some of the closely related recent works on contact-less sleep monitoring, which include some sound, radar and WiFi CSI based techniques.

Lullaby~\cite{kay2012lullaby} tracks various environmental factors, sound, light, temperature, and motion that help users assess the quality of their sleep environments.
iSleep \cite{hao2013isleep} uses the built-in microphone of the smartphone to detect the events that are closely related to sleep quality, including body movement, couch and snore, and infers quantitative measures of sleep quality. %(Data: 7 participants and total 51 nights of sleep).
Sleep Hunter \cite{gu2014intelligent} uses actigraphy and acoustic events to predict sleep stage transitions by smartphone.
Toss-N-Turn \cite{min2014toss} uses features such as sound amplitude, acceleration, light intensity,
screen proximity, battery and screen states, etc. to track a subject's sleep quality. 
However, Audio based techniques are privacy invasive, and therefore, often avoided as sleep is a private activity.

RF sensing based techniques are by far the least intrusive methods for monitoring sleep, both in terms of privacy and convenience of use.
DoppleSleep \cite{rahman2015dopplesleep} is another unobtrusive sleep sensing system which uses short-range Doppler radar to perform sleep stage classification (Sleep vs. Wake and REM vs. Non-REM). 
Vital-radio \cite{adib2015smart} develop an FMCW based system which is shown to accurately track a person's breathing and heart rate without body contact, from distances up to 8 meters.
Based on the same system, \cite{zhao2017learning} proposes a deep learning architecture to perform 4-stage sleep stage classification.
More recently, authors of \cite{yue2018extracting} proposed algorithms to achieve multi-person identification and breath monitoring based on the same FMCW hardware. 
Although the aforementioned radar based techniques do fairly well in terms of monitoring vital signs during sleep. However, they require dedicated hardware and spectrum, adding cost, scalability\footnote{e.g., when multiple such radars are co-located}, and/or RF regulation hurdles.
These factors prevent their large-scale and long-term deployment.
WiFi signals based sensing has recently emerged an approach to low-cost and easily adoptable long-term sleep monitoring, as the widespread use of WiFi capable devices (\eg smart-home assistants, smart-phones, etc.) has made WiFi signals the most ubiquitous form of sensing in homes requiring no additional hardware costs.
Multiple WiFi CSI based schemes have been proposed for tracking vital signs during sleep \cite{hillyard2018experience, liu2015tracking, liu2014wi, wang2016human, zhang2018fresnel}.
The key limitation of the existing works is that they have only been evaluated with short-duration mock sleep experiments in very controlled settings.
This makes the applicability of their techniques and findings quite limited in practical sleep monitoring scenarios.
For example, the techniques proposed in existing works have been designed based on controlled experiments often performed on the same subject, where they require the subject to lie down in between and/or very close to both transmitter (TX) and the receiver (RX) to ensure line-of-sight (LOS) scenarios.
Their techniques rely on trial-and-error based positioning of WiFi transceivers and signal processing methods to track vital signs, which often leads to high dependency on multiple, environment-dependent parameters that are difficult to tune in real world.
Such techniques may be suitable for controlled short-duration lab experiments.
However, their suitability cannot be generalized to different individuals, environments, positioning of WiFi transceivers, LOS/NLOS situations, and to natural in-home full-night sleeping scenarios.

 \presec
\section{Understanding The Relationship between Vital Signs and WiFi CSI} \label{sec:workingprinciples}
\postsec

\subsection{Overview of WiFi CSI} \label{sec:csioverview}
\postsub

WiFi devices measure the Channel State Information (CSI), which characterises the surrounding wireless channel across bandwidth and multiple antennae.
The Orthogonal Frequency Division Multiplexing (OFDM) communication scheme used in IEEE 802.11a/n/ac divides the wireless channel's bandwidth into multiple modulated subcarriers.
To correct for channel frequency-selectivity (or equivalently the \textit{delay spread} in time-domain) and maximise the link's capacity, WiFi devices continuously track changes over these subcarriers in terms of CSI values, which are then used to adapt transmission power and rates in real time.
CSI values are the \textit{Channel Frequency Response} (CFR) at per subcarrier granularity between each transmit-receive (Tx-Rx) antenna pair.
When a user is breathing, the chest and body movements change the constructive and destructive interference patterns of the WiFi signals.
The CSI values are sensitive enough to measure these breathing movements, as CSI measurements can be obtained at high sampling rates and from multiple different OFDM subcarriers of each TX-RX stream.
For example, the driver of the Intel 5300 WiFi NIC, which we use to implement our scheme, reports CSI values on $30$ OFDM subcarriers \cite{halperin2011tool} for each TX-RX antenna pair for every CSI measurement.
This leads to $30$ matrices with dimensions $M_t \times M_r$ per CSI sample, where $M_{t}$ and $M_{r}$ denote the number of transmit and receive antennas respectively.
Such high dimensional data allows us to recover detailed information about the vital signs even if the breathing and body/limb related movements only incur small changes in the CSI.

\presub
\subsection{Impact of Breathing on WiFi Multipath} \label{sec:breathmultipath}
\postsub
Next, we present our first analysis that is aimed at understanding the effect of small breathing movements on the WiFi multipath and CSI signals.
Based on this analysis, we design Serene's signal processing pipeline to robustly extract breathing waveforms in an individual and environment independent manner.
Our analysis shows that if we differentiate (\ie by taking \textit{first order difference}) the CSI signals from each WiFi subcarrier, and then \textit{max-min normalize} the CSI signal projection corresponding to variations due to breathing, we can robustly extract the waveform corresponding to user's breathing motion in an environment and individual independent manner,
as long as the user sleeps close to the WiFi receiver.
Such proximity requirement is easy to satisfy during real-life in-home sleep scenarios by either mounting receiver on the headboard of a bed frame or placing it on a table nearby. 
\textit{Note that differentiation generally degrades signal-to-noise ratio, unless the differentiation algorithm includes smoothing that is carefully optimized for the application at hand. 
Therefore, we introduce a combination of low-pass filters (i.e. median, exponential moving average, and Butterworth filters (\S \ref{sec:noiseremoval_0})) in Serene's signal processing pipeline.
We experimentally design these filters such that signal-to-noise ratio is sufficiently good for a reasonable quantitative measurement of the sleep related vital signs.
The max-min normalization is performed after the filtering process to estimate the breath rate (\S\ref{sec:breathratemeasurement}).} 
At the basis of our analysis is a closed form expression, which we derive using time-varying Channel Frequency Response (CFR) of WiFi channel.
The time-varying CFR corresponding to a Tx-Rx antenna pair for a subcarrier with wavelength $\lambda$ can be quantified as:

\vspace{-0.35cm}
\begin{equation} \label{eqn:1}
    H(f,t) = H_s(f) + \underbrace{\sum_{i=1}^{N} \frac{K}{D_i(t)^2} \: e^{\frac{j2\pi D_i(t)}{\lambda}}}_{H_d(f,t)}
    \vspace{-0.02in}
\end{equation}

In the equation above, $N$ is the number of 
multipath reflections of the transmitted signal at the Rx end, $D_i$ represents the distance traveled by $i^{th}$ multipath reflection,  and $K$ is an environment dependent proportionality constant.
$H_s(f)$ is the \textit{static} component of CFR corresponding to all non-user multipath reflections, while the second term on the right hand side corresponds to the \textit{dynamic} component of CFR, represented as $H_d(f,t)$, while the user is breathing and/or moving during sleeping.
Now, let us assume that user is sleeping at a distance $D_{0,i}$ from the router, and $d_i(t)$ is the change in distance traveled by $i^{th}$ reflected path due to breathing. 
To make our scheme resistant to static changes in the environment, we first eliminate $H_s(f)$ by differentiating the above equation with respect to $t$, and substitute $D_i(t) = D_{0,i} + d_i(t)$ to get:

\vspace{-0.12in}
\begin{equation} \label{eqn:2}
\begin{split}
H'(f,t) = \frac{d}{dt} \Bigg[\sum_{i=1}^{N} \frac{k}{D^2_{0,i}} \Big(1+\frac{d_i(t)}{D_{0,i}}\Big)^{-2} e^{\frac{j2\pi (D_{0,i} + d_i(t))}{\lambda}}\Bigg]
\vspace{-0.05in}
\end{split}
\end{equation}

As $d_i(t)$ caused by motion due to breathing is in the order of a few centimeters, whereas $D_{0,i}$ is usually in the order of meters (i.e. $d_i(t) \ll D_{0,i}$), we can expand the negative polynomial $(1 + \frac{d_i(t)}{D_{0,i}})^{-2}$ via binomial series expansion.
After performing binomial expansion, discarding the $\frac{d_i(t)^m}{(D_{0,i})^n}$ terms with n = 4 or higher, and doing some algebraic manipulations, we get the following expression for $H'(f,t)$:

\vspace{-0.1in}
\begin{equation*} \label{eqn:4}
\begin{split}
H' \approx ke^{\frac{j2\pi D_{0,i}}{\lambda}} \times \Bigg[\sum_{i=1}^{N} d'_i(t) \Bigg(-\frac{2}{D^3_{0,i}} + \\  j\Bigg( \frac{2\pi}{\lambda D^2_{0,i}} - \frac{4\pi d_i(t)}{\lambda D^3_{0,i}}\Bigg) \Bigg) e^{\frac{j2\pi d_i(t)}{\lambda}}\Bigg]
\end{split}
\end{equation*}

After converting the term inside summation into polar coordinates, and discarding the $\frac{d_i(t)^m}{(D_{0,i})^n}$ terms with n = 4 or higher, we get the following simplified expression for $H'(f,t)$:

\vspace{-0.05in}
\begin{equation*} \label{eqn:5}
\begin{split}
H' \approx \Bigg[\frac{2\pi k}{\lambda D^2_{0,i}} \sqrt{1 + \Bigg( \frac{\lambda}{\pi D_{0,i}}\Bigg)^2 } \times \\ e^{\frac{j2\pi D_{0,i}}{\lambda}}\Bigg] \cdot \Bigg[ \sum_{i=1}^{N} d'_i(t) e^{\frac{j2\pi d_i(t)}{\lambda} + jA_i} \Bigg]
\end{split}
\end{equation*}

\begin{figure}[htbp]
    \centering
    \captionsetup{justification=centering}
    \subfigure[Variation of $A_i$ with $d_i(t)$ for $D_{0,i}$ 3 to 10]{
        \includegraphics[width=0.335\textwidth]{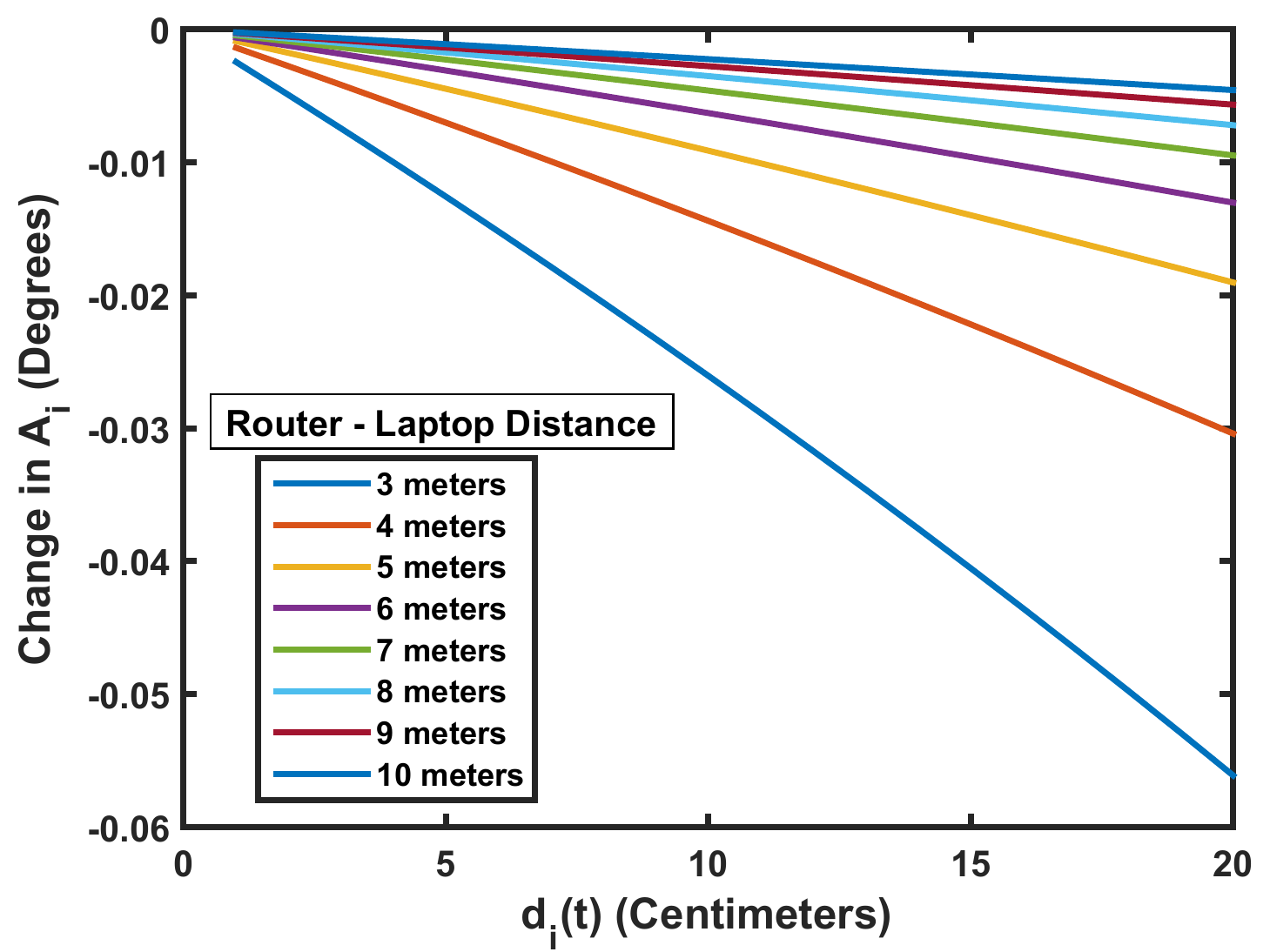}
        \label{fig:ChangeInTanInverse}
    }
    \subfigure[Single breath samples for 7 configurations]{
        \includegraphics[width=0.335\textwidth]{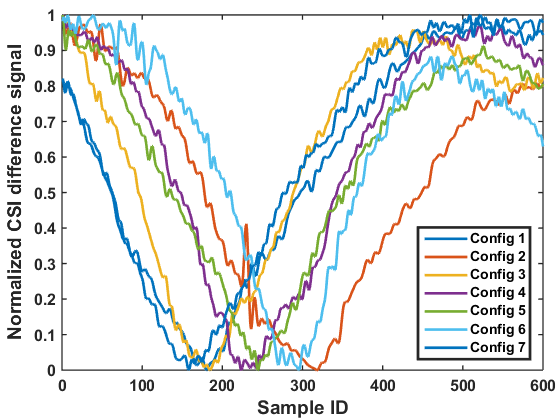}
        \label{fig:examplesample}
    }
    \subfigure[Configurations of receiver]{
        \includegraphics[width=0.275\textwidth]{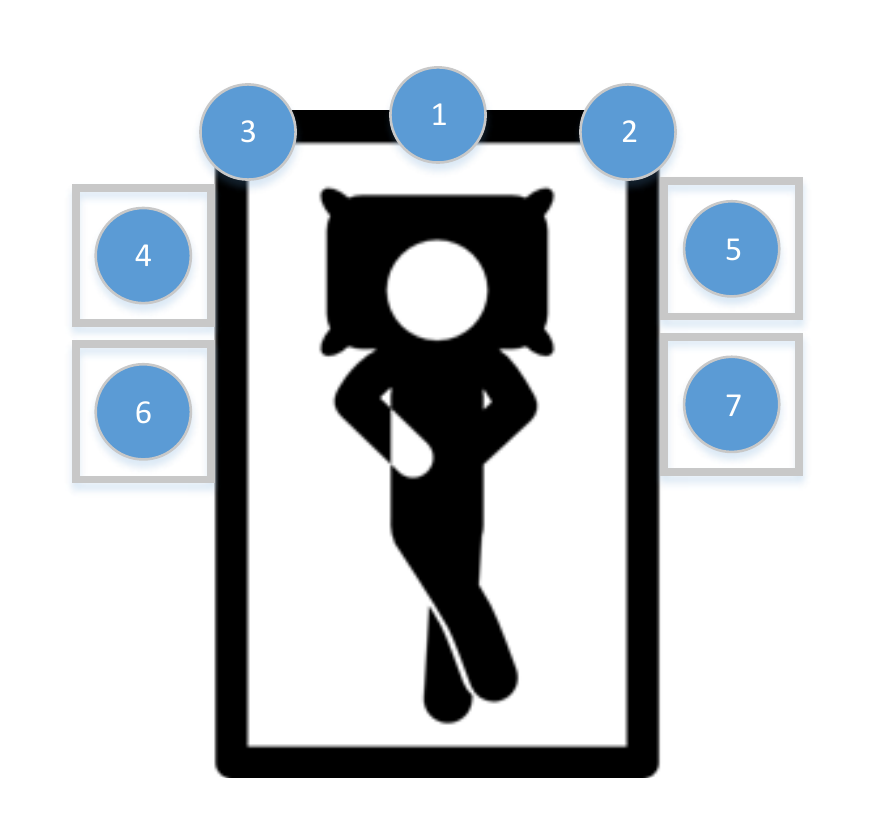}
        \label{fig:examplesamplepic}
    }
    \vspace{-0.15in}
    \caption{(a) Variation of $A_i$ with $d_i(t)$ for different $D_{0,i}$; (b) Single breath samples for different configurations in (c)}
    \label{fig:examplesamplewithpic}
    \vspace{-0.14in}
\end{figure}

Here, $A_i = tan^{-1}\Big[\frac{\pi D_{0,i}}{\lambda} \Big(1 - \frac{2\cdot d_i(t)}{D_{0,i}}\Big) \Big]$.
Figure \ref{fig:ChangeInTanInverse} shows variation of $A_i$ with $d_i(t)$, as $d_i(t)$ varies from 1cm to 20cm (typical range for motion due to human breathing is 1-5cm \cite{long2015analysis}), for different router-receiver distances $D_{0,i}$ ranging from 3m - 10m (\ie which is typical for regular home use cases).
We observe that changes in $d_i(t)$ do not significantly affect the value of $A_i$. Moreover, the impact of $d_i(t)$ on $A_i$ decreases even further as the distance between receiver and the router it is connected to increases.
Therefore, we can safely approximate $A_i \approx tan^{-1}\Big[\frac{\pi D_{0,i}}{\lambda} \Big] = A_{0,i}$ and write $H'(f,t)$ as:

\vspace{-0.05in}
\begin{equation*} \label{eqn:6}
\begin{split}
H'\approx\Bigg[\Bigg(\frac{2\pi k}{\lambda D^2_{0,i}} \sqrt{1 + \Bigg( \frac{\lambda}{\pi D_{0,i}}\Bigg)^2 } \Bigg) \times \\ e^{j\Big(\frac{2\pi D_{0,i}}{\lambda} + A_{0,i}\Big)}\Bigg] \times \sum_{i=1}^{N} d'_i(t) e^{\frac{j2\pi d_i(t)}{\lambda}}
\end{split}
\end{equation*}

The first term on right hand side of the equation above stays constant when receiver is placed on some surface, \eg a desk/table, and is not moving. We write amplitude of CFR \ie $|H'(f,t)|$ as: 

\vspace{-0.12in}
\begin{equation} \label{eqn:7}
|H'(f,t)| \approx C_{0,i} \cdot \Big|\sum_{i=1}^{N} d'_i(t) e^{\frac{j2\pi d_i(t)}{\lambda}}\Big|
\end{equation}

The waveform $\Big|\sum_{i=1}^{N} d'_i(t) e^{\frac{j2\pi d_i(t)}{\lambda}}\Big|$ corresponds to the variations due to breathing. The proportionality term $C_{0,i}$ in breathing samples extracted from $|H'(f,t)|$ corresponding to different placement of receiver can be easily eliminated via \textit{max-min normalization}.
Figure \ref{fig:examplesample} shows extracted and processed single breath samples from a user for seven slightly different receiver placement configurations close to the user, while the router was in subject's TV-lounge (router-receiver distance \textgreater 10 meters).

\presub
\subsection{Impact of Breathing and Body/Limb Movements on WiFi Signal Subspace} \label{sec:breathsubpace}
\postsub
Next, we present our second analysis that is aimed at understand how breathing affects the signal subspace formed by WiFi subcarriers compared to other bodily movements.
Today's MIMO and OFDM based WiFi devices use many frequency subcarriers and multiple transmit-receive (Tx-Rx) antennas for data communication. 
The MIMO system between the OFDM subcarriers and the Tx-Rx antennas, forms a multidimensional array which effectively represents a high-dimensional mathematical space. 
Contained in this space is the signal subspace along frequency and spatial dimensions~\cite{Alloulah18_SubspaceTrackingWifi}.
The key intuition behind our model is that while a user is sleeping, the signal subspace along these dimensions is affected by both breathing and body/limb motion. 
When there is no body/limb motion, there is only one dominant time-varying component in the subspace, which corresponds to breathing. 
However, more components along these dimensions evolve (\ie show considerable variations) during other body/limb activity \eg during roll overs or arm/leg movement.
Based on this principle, Serene isolates breathing from limb motion without requiring any environment-dependent calibrations.

\begin{figure*}[htbp]
    
    \captionsetup{justification=centering}
    \begin{center}
        \captionsetup{justification=centering}
        \subfigure[Power values in top PCA projections]{
            \includegraphics[width=0.35\textwidth]{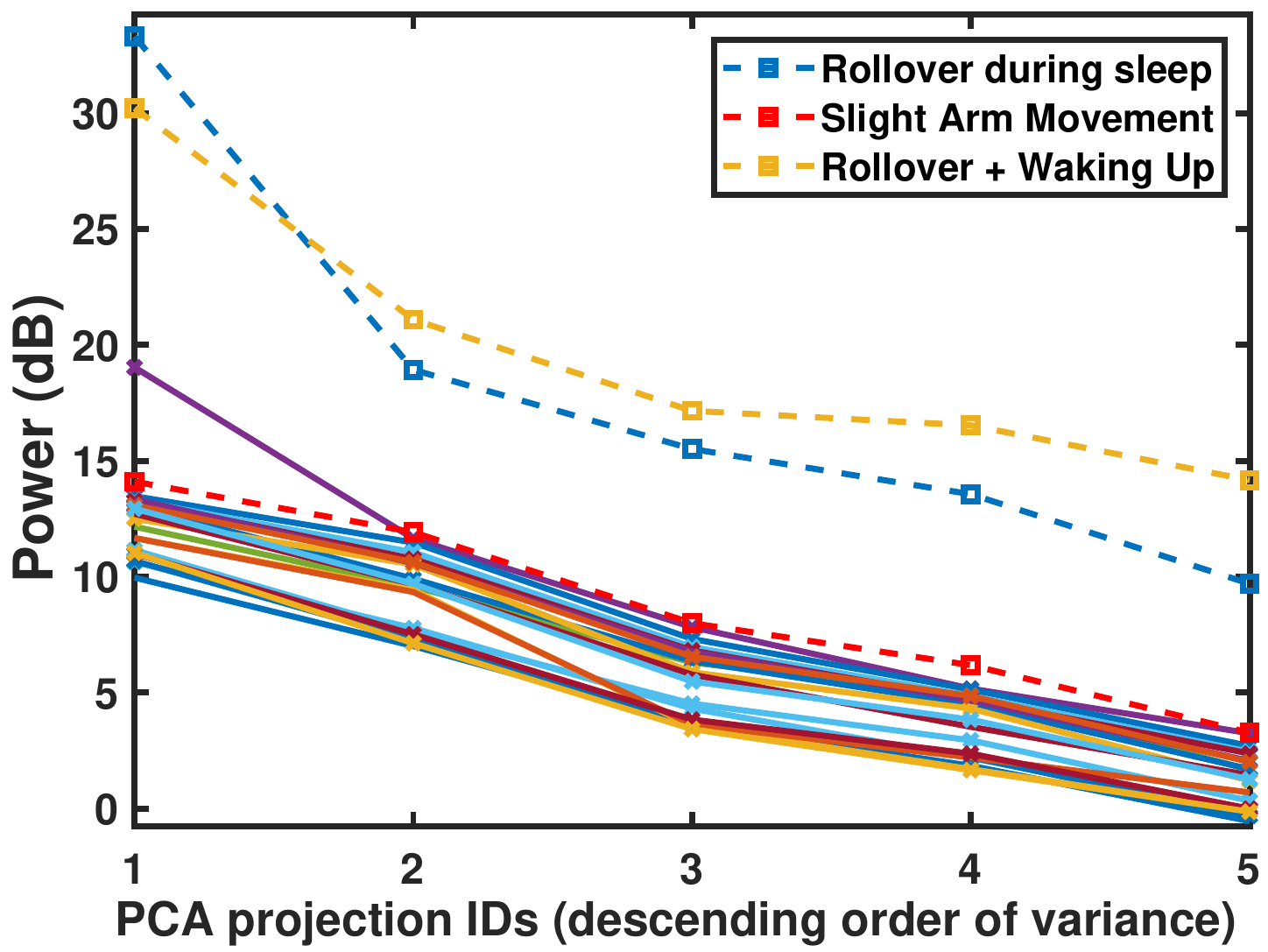}
            \label{fig:breath-subspace-model-1}
        }
        \vspace{-0.1in}
        \subfigure[CSI timeseries for Fig. \ref{fig:breath-subspace-model-1}]{
            \includegraphics[width=0.35\textwidth]{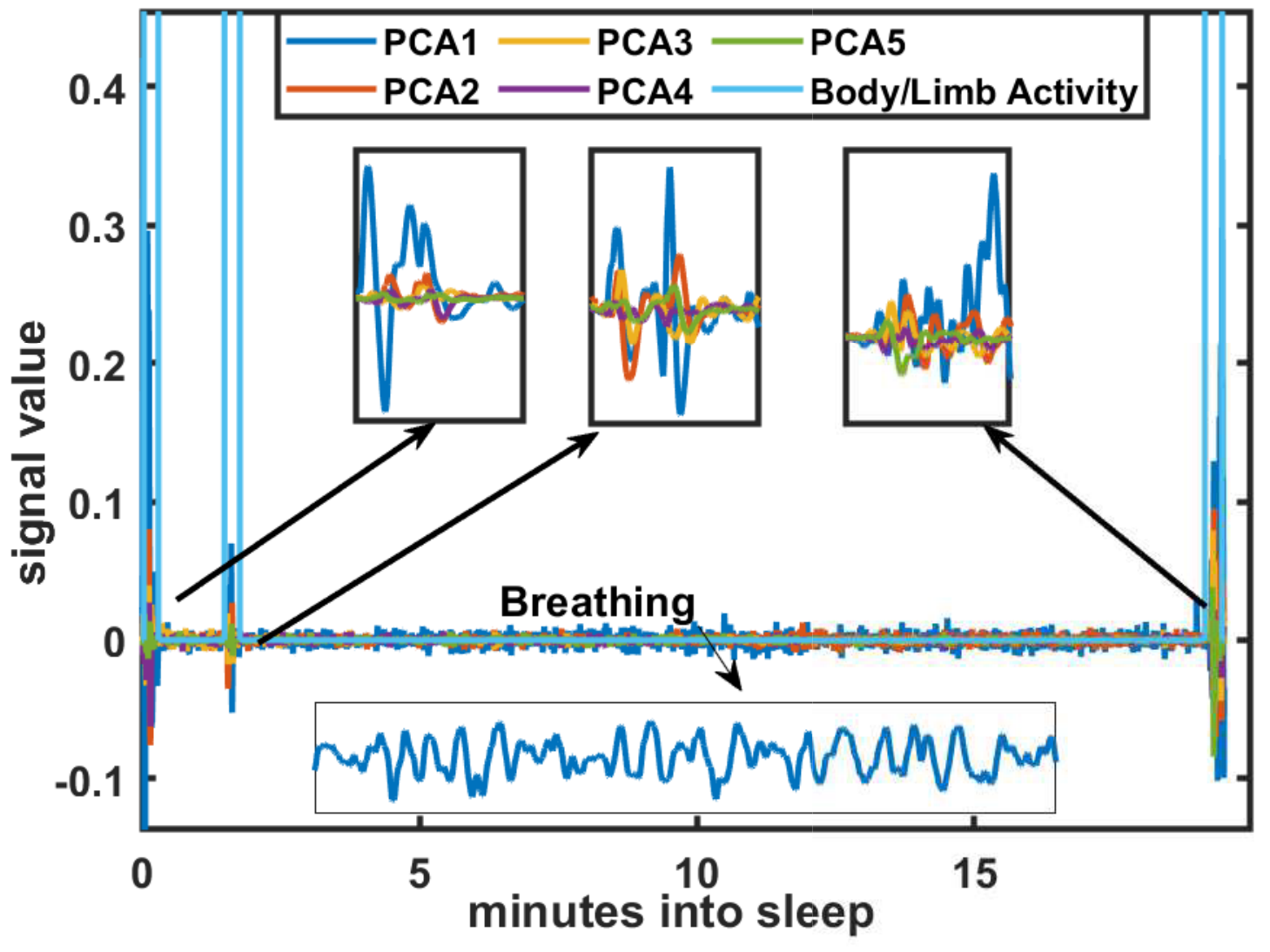}
            \label{fig:breath-subspace-model-2}
        }
    \end{center}
    \vspace{-0.14in}
    \caption{Impact of bodily activity during sleep on WiFi subspace}
    \label{fig:breath-subspace-model}
    \vspace{-0.14in}
\end{figure*}

To model this in Serene, we track the top dominant components in the CSI signal subspace using Principal Component Analysis (PCA).
Figure \ref{fig:breath-subspace-model-1} shows power values in top 5 PCA projections of the CSI signals corresponding to multiple sleep epochs during a sleep experiment,
where the dotted lines correspond to epochs with motion events---highlighted in Figure~\ref{fig:breath-subspace-model-2}.
%   %
We observe that in the absence of any body/limb activity, the top-most PCA projection is enough to track breathing as it is the only major motion occurring in the environment. 
%   %
However, during body/limb movements, multiple lower PCA projections also show significant variations. 
%   %
Based on this phenomena, we accurately detect and then discard the CSI values corresponding to any body/limb activity by tracking variations in the lower PCA components (\eg 3, 4 and 5) using a multi-dimensional clustering technique, which we discuss in \S \ref{sec:featureextraction}.

 \begin{figure*}[htbp]
    \centering
    \captionsetup{justification=centering}
    \includegraphics[width=0.95\textwidth]{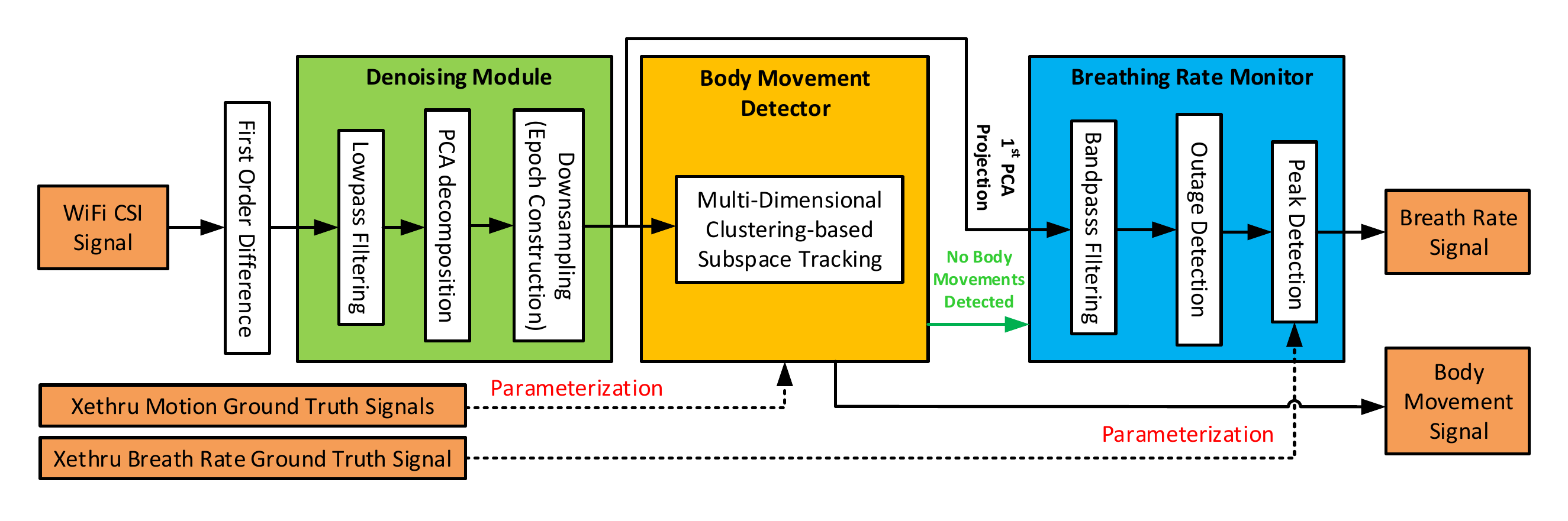}
    \vspace{-0.22in}
    \caption{Our WiFi CSI signal processing architecture for extracting vital signs}
    \label{fig:system}
    \vspace{-0.2in}
\end{figure*}

\presub
\section{CSI Signal Processing Architecture} \label{sec:noiseremoval}
\postsub

To obtain CSI data in real-time during sleep, we develop a client-server based mechanism to communicate the CSI values extracted from WiFi NIC to a Python based CSI processing server.
Based on our discussion in \S \ref{sec:workingprinciples}, we take first order difference of the incoming CSI data and then take its amplitude \ie $|H'(f,t)|$ for further processing.
From now onward, we use the term ``CSI'' to denote $|H'(f,t)|$.
CSI data is collected in 30 second epochs, which is typically the partitioning convention followed by most sleep monitoring systems.
Next, we first perform basic low-pass filtering for removing bursty noise due to hardware noise and isolate the signal of interest i.e. to extract human motion related frequencies only.
Second, we perform PCA on the low-pass filtered CSI streams for dimensionality reduction and automatic distinguishing of CSI variations due to body movements from those of breathing in different subcarriers based on our discussion in \S \ref{sec:breathsubpace}.
This avoids the need for complex trial-and-error based subcarrier selection procedures used in previous works \cite{liu2015tracking, liu2014wi}. 
Third, we harmonise the filtered CSI data corresponding to each 30s sleep epoch into uniformly sampled and consistent measurements via down-sampling.
Fourth, we robustly detect body movements by tracking lower PCA projections of CSI signals using a clustering-based event detection technique.
Finally, we first detect the presence of breathing using a power threshold, and then perform further band-pass filtering to extract the signal corresponding to breathing.
Figure \ref{fig:system} shows our system architecture.
Next, we briefly discuss Serene's noise removal process.

\presub
\subsection{Noise Removal} \label{sec:noiseremoval_0}
\postsub

Commodity Wi-Fi NICs report noisy CSI values, both due to hardware limitations (such as low resolution Analog to Digital Converters (ADCs)) and due to changing transmission power and rates.
We use a combination of median filter and an exponential moving average filter to get rid of such bursty noise and spikes in CSI time series.
After this basic filtering step, we further remove any high frequency variations in CSI signals using \textit{Butterworth} low-pass filter. 
Variations due to movement during sleep cause low frequency variations, typically under 5 Hz \cite{long2015analysis}.
We use Butterworth low-pass filter for separating these variations from higher frequency noise in CSI values. Due to maximally flat amplitude response of Butterworth filter, its application on CSI time series does not distort the shape of CSI variations due to body motion.
Our scheme samples CSI values at a nominal frequency $F_{s} = 800$. With this in mind, we use cut-off frequency $\omega_{c} = \frac{2\pi * f}{F_{s}} = 0.0125\pi$ rad/s for Butterworth filtering.
We apply the same filter on CSI timesseries of all the subcarriers, making sure that every CSI stream experiences the same phase distortion and group delay introduced by the filter.
Based on our experimental results, we observed that filtered CSI waveforms still retain some noisy variations which are not related to activity/breathing. 
We avoid any further low pass filtering on CSI streams as it can lead to loss in details of variations due to activity/breathing behavior.
To remove such noise, we utilize the fact that the CSI variations in CSI streams of multiple subcarriers in each Tx-Rx antenna pair are correlated.
We apply PCA on CSI obtained from all subcarriers and all Tx-Rx pairs, and retain only the waveforms that represent the most common variations in all the subcarriers, i.e., the variations due to breathing and/or body movements during sleep. 
That is, signal subspace-based filtering enables our scheme to automatically obtain the signals that are representative of the monitored vital signs only. 
PCA also reduces the dimensionality of data by discarding the principal components unrelated to the vital signs i.e. the noise subspace.
Finally, we rearrange the multi-dimensional filtered CSI data corresponding to each 30s sleep epoch into consistent length samples (600 in our current implementation) via down-sampling, performing zero-padding where necessary.
Although we have partitioned the incoming CSI data into 30s epochs, we concatenate data from consecutive epochs for real-time tracking of vital signs (\eg breathing) which we discuss later in this section.

 \presub  
\subsection{Tracking Body Movements} \label{sec:featureextraction}
\postsub  

As discussed earlier, today's MIMO and OFDM based WiFi devices use multiple frequency subcarriers and Tx-Rx antennas for data communication. 
The MIMO system between the Tx-Rx antennas, and the OFDM subcarriers, form a multidimensional tensor along space-frequency axes.
Contained in such tensor is the signal subspace we wish to track for breathing and body motion. 
We observed that when there is no body/limb motion, there is only one dominant, time-varying component in the signal subspace, which corresponds to breathing. 
PCA sorts different principal components in descending order of their variation. 
During sleep, the signal subspace is rather quiet and breath is captured in the top PCA projection of the CSI time series.
However, we observed that during episodes of other body movements---\eg during roll overs or arm/leg movement---more signal subspace components evolve, since body movements cause more pronounced variations in the spatial-frequency subspace compared to faint breathing movements. 

\subsubsection{Body Movements Detection Approach}
To robustly distinguish body activity/limb motion from breathing, we propose to use a multi-dimensional \textit{hyper-ellipsoidal} clustering on the lower PCA projections of CSI data.
At a high-level, we can think of this clustering method as a high-dimensional generalization to a Gaussian outlier rejection technique whereby measurements few sigma's away from the mean are deemed erroneous.
Specifically, let $R_{k}=\{r_{1},r_{2},\cdots r_{k}\}$ be the first $k$ samples of CSI vectors containing values from the selected signal subspace---\textit{we use PCA projections 3, 4 and 5 in our current implementation}.
Each sample $r_{i}$ is a $d\times 1$ vector in $\mathbb{R}^{d}$, where $d$ is the number of signal subspace components. 
This \textit{hyper-ellipsoidal} technique clusters the normal data points (i.e. when there is no body movement in the environment), and any points lying outside the cluster are declared as outliers. 
The boundary of the cluster (a ``hyper-ellipsoid' in this case) is related to a distance metric which is a function of mean $m_{R,k}$ and covariance $S_{k}$ of the incoming signal subspace components $R_{k}$.  
We use the \textit{Mahalanobis} distance metric, $D_i$, for which the cluster is arrived at according to the following ~\cite{Mosht2011}:

\vspace{-0.1in}
\begin{equation}\label{ellipse}\small
    e_{k}(m_{R},S_{k}^{-1},t) =\{r_{i}\epsilon \mathbb{R}^{d}|  \underbrace{\sqrt{(r_{i}-m_{R,k})^{T}{S_{k}}^{-1}(r_{i}-m_{R,k})}}_{D_{i}=Mahalanobis \hspace{2mm}distance\hspace{2mm}of\hspace{2mm}r_{i}}\leq t\}
\end{equation}

where $e_k$ is the set of normal data points  whose Mahalanobis distance, $D_i <t$ and $t$ is the \textit{effective} radius of the hyper-ellipsoid. The choice of $t$ depends on the distribution of the normal data points. 
If the normal data follows a \textit{chi-squared} distribution, it has been shown that up to 98\% of the incoming normal data can be enclosed by the boundary of an hyper-ellipsoid, if the effective radius $t$ is chosen such that $t^{2}=({\chi_{d}}^{2})_{0.98}^{-1}$ \cite{Mosht2011}.  
Data samples $r_{i}$ for which $D_i > t$, are therefore, identified as outliers. As it is often not practical to store all the samples of a streaming data, therefore a recursive algorithm is required to update $e_{k}$. Let $r_{k+1}$ be the most recent CSI sample. $m_{R,k+1} = \frac{km_{R,k} + r_{k+1}}{k+1}$ and $m_{R^2,k+1} = \frac{km_{R^{2},k} + r_{k+1}r_{k+1}^{T}} {k+1}$ can be updated recursively from the previous means.
By substituting covariance matrix $S_{k} = m_{R^{2},k} - (m_{R,k} \: m_{R,k}^{T})$ into Eq.~\eqref{ellipse} we can represent $e_{k}$ entirely in terms of means. The resulting equation updates the cluster boundary and classifies the incoming data samples as normal readings or outliers. Our scheme uses above equations for initial estimation of mean and covariance. Afterwards, the mean $m_{R,k}$ is recursively calculated using an exponential moving average technique, where mean $m_{R,k+1}$ is updated as $m_{R,k+1} = \alpha ~ m_{R,k} + (1-\alpha) r_{k+1}$,
where $\alpha = 0.9995$ in our implementation. Moreover, after initial estimation of covariance, our scheme recursively updates the covariance inverse $S_{k}^{-1}$ by using the following equation \cite{Mosht2011}, which avoids extra computations required for calculating the inverse of matrix $S$:

\vspace{-0.05in}
\begin{equation}\label{cov1}
\begin{split}
    S_{k+1}^{-1} = \frac{kS_{k}^{-1}}{\alpha (k-1)}\times \hspace{5cm} \\ \left[I - \frac{(r_{k+1} - m_{R,k})(r_{k+1} - m_{R,k})^{T}S_{k}^{-1}}{\frac{(k-1)}{\alpha} + (r_{k+1} - m_{R,k})^{T}S_{k}^{-1}(r_{k+1} - m_{R,k})}\right]
\end{split}
\end{equation}

To determine the start and end of activity waveforms, we use both cardinality and temporal proximity of the detected outliers. If the number of consecutive outliers increases a threshold $E_1$, we declare a micro-event. Multiple micro-events constitute a activity event. All the data points including the points constituting the micro-events as well as the points in between the consecutive micro-events are recorded as part of activity waveform (\textit{merging}). We only merge the micro-events which are within $E_2$ data points of each other.
Both E1 and E2 are design time, easy to tune thresholds.
Figure \ref{fig:movementdetectiondetail} shows some body movements detected by our algorithm in a portion of processed CSI timeseries corresponding to an in-home full-night sleep monitoring experiment.

\begin{figure*}[htbp]
    \vspace{-0.125in}
    \centering
    \captionsetup{justification=centering}
    \includegraphics[width=1\textwidth]{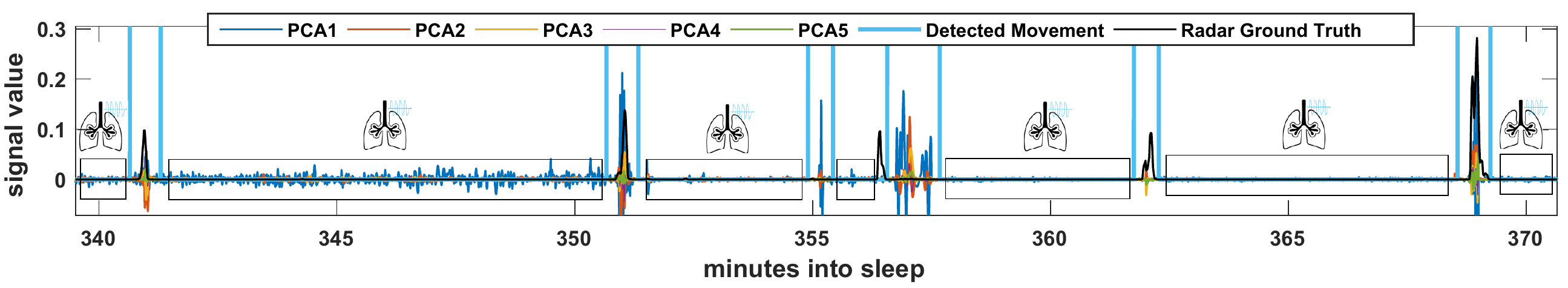}
    \vspace{-0.24in}
    \caption{Example showing performance of our body movement detection algorithm, compared to Xethru radar ground truth. Boxes show the areas where breathing is usually present. Ground truth is approximately synchronized with CSI data.}
    \label{fig:movementdetectiondetail}
    \vspace{-0.2in}
\end{figure*}

\presec
\subsection{Tracking Breath} \label{sec:tracking_respiration}

Human breath involves motion of chest and lungs during inspiration (when air enters the lungs) and expiration (when air is blown out from the lungs) \cite{long2015analysis}.
These motions are often periodic (e.g. in case of healthy subjects \cite{long2015analysis}), and therefore, cause periodic variations in WiFi channel which we can extract using CSI data.
In the absence other body movements, the first PCA projection of CSI data would be able to capture these variations due to breathing.
However, as these minute variations are often embedded in noise, and because human subject might not be in proximity of the RX device, we can not always assume that breathing signal exists in a particular sleep epoch.
Therefore, to robustly track breathing, we propose the following signal processing pipeline.

\subsubsection{Bandpass Filtering}
To extract the periodic variations in CSI due to breathing, we apply a Butterworth bandpass filter on the first PCA projection.
We choose the filtering parameters of this filter according to the fact that breathing rate of humans (including adults as well as newly born babies) ranges between 10 - 40 breaths per minute (BPM) \cite{long2015analysis}. 
This step removes any non-breathing related noise present in the signal.

\subsubsection{Measuring Breathing Rate} \label{sec:breathratemeasurement}
We design our system to measure breathing rate in terms of breaths per minute (BPM).
We measure the rate over a window of two sleep epochs in length, which moves over concatenated data from multiple consecutive sleep epochs.
To report BPM every second, we move this window over the concatenated data every second (i.e. 20 samples a time).

To measure breathing rate, we employ a \textit{peak detection} based approach.
First, we max-min normalize the signal so that parametrization of our peak detection algorithm can be easily generalized to different users.
Second, to robustly detect the number of peaks, we use three parameters, namely \textit{minimum peak prominence} (MINPRO), \textit{minimum peak distance} (MINDIST), and \textit{minimum peak strength} (MINSTR).
The prominence of a peak measures how much the peak stands out, due to its height and location, relative to other peaks around it. 
We tune MINPRO such that we only detect those peaks which have a relative importance of at least MINPRO.
We tune MINDIST according to the fact that human breathing rate ranges between 10-40 BPM \cite{long2015analysis}, so that redundant peaks are discarded. 
To further sift out redundant peaks, we only choose peaks of value greater than MINSTR times the median peak value.
In our current implementation, we chose MINPRO = 0.025, MINDIST = 1.5 seconds and MINSTR = 0.6 which generalize well for different sleeping scenarios.
To achieve accurate tracking of breathing rate, we perform parameterization of Serene's breath estimation algorithm using ground truths obtained from a contact-less COTS Xethru X4M200 Breath/Motion sensor \cite{xethru200}.
We perform this parametrization only during the design of our system and do not require any end-user calibration effort in the real-world deployments.

 \presec
\section{Implementation and Evaluation}
\postsec
In this section, we present the performance evaluation of our system after performing full-night sleep experiments in real-world settings.
Next, we first discuss our hardware implementation and the experimental settings.

\begin{figure*}[htbp]
    \centering
    \captionsetup{justification=centering}
    \includegraphics[width=0.8\textwidth]{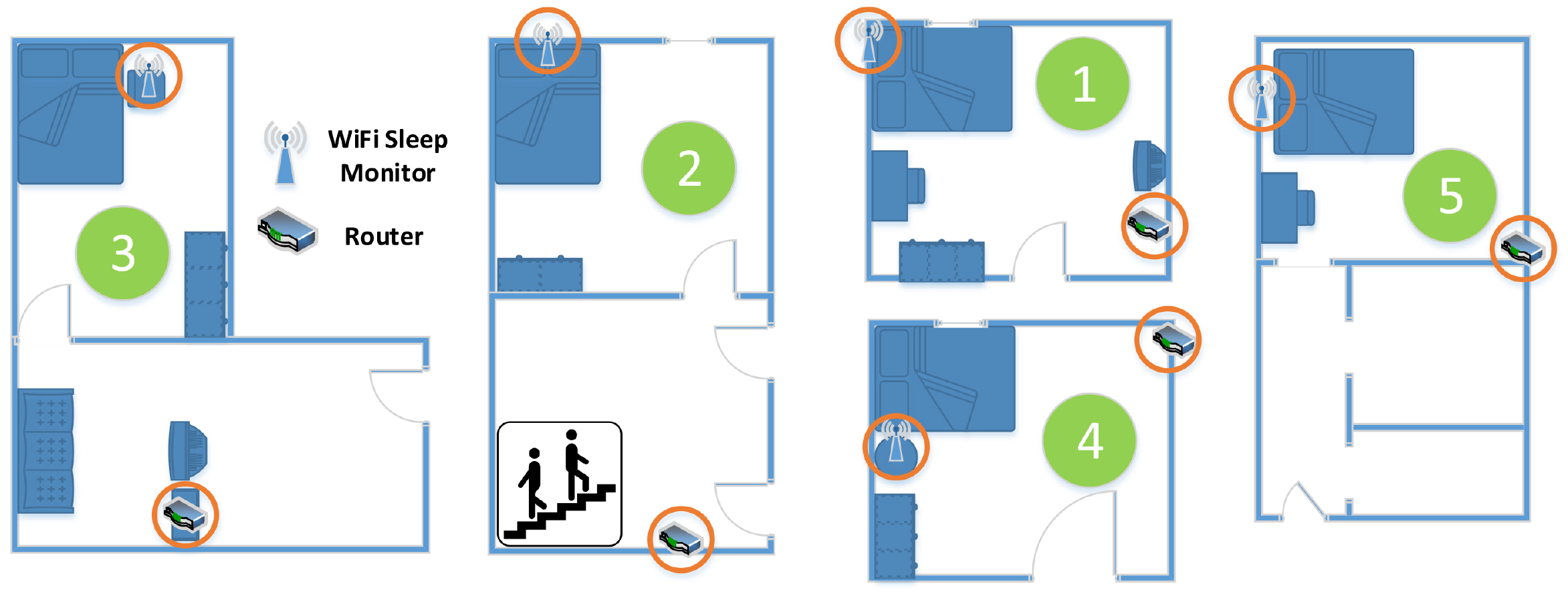}
    \vspace{-0.1in}
    \caption{The real-world deployment scenarios used for evaluation of our sleep monitoring scheme ``Serene''.}
    \label{fig:experimentalsetups}
    \vspace{-0.15in}
\end{figure*}

\presec
\subsection{Hardware Implementation}
\postsec
We developed compact HummingBoard (HMB)-based small-sized nodes as sleep monitoring devices~\cite{hummingboard} which makes Serene easy to deploy.
HMB nodes were equipped with the Intel 5300 NICs with modified drivers for extracting CSI information \cite{halperin2011tool}. 
We used Linksys AC1200+ routers as transmitters in our deployments.
Moreover, we developed a client-server software architecture---in C and Python respectively---capable of the real-time extraction and processing of CSI data throughout the night.
For body movement and breathing rate ground truths, we deployed state-of-the-art \textit{pulse-doppler} radar-based Xethru X4M200 Breath/Motion sensors \cite{xethru200}. 
In terms of breathing rate accuracy, the X4M200 devices have been shown to perform very closely to a medical-grade, gold standard equipment (X4M200 has been shown to track breathing with up to 96\% accuracy when compared to PSG) \cite{xethrureliability}.
We chose a contact-less sensor to record ground truths because the participants of our study were not comfortable wearing devices such as breath monitoring belts during their regular sleep.
Moreover, the devices that require body contact generally tend to interfere with natural sleep of the users \cite{choe2010opportunities}.
We utilize these ground truths in our system for: (1) the robust parametrization of our signal processing pipelines (e.g. breath tracking), (2) measuring breath tracking accuracies, and (3) measuring limb/body motion tracking accuracies.

\presec
\subsection{Experimental Settings}
\postsec
We deployed our system in 5 apartments, where we collectively recorded more than 80 nights (>550 hours) of data from 5 different participants.
The participants were graduate students aged between 23 to 32 years.
The duration of data collection for each participant varied from 5 to 31 days.
Figure \ref{fig:experimentalsetups} shows the real-world deployment scenarios for our sleep experiments.
The numbers in circles specify user/environment IDs.
Data collected from environments 2 and 3 corresponds to NLOS deployment scenario, and constitute ~55\% of our dataset.   
Data collected from environments 1, 4, and 5 corresponds to LOS deployment scenario, and constitute ~45\% of our dataset.
To evaluate Serene's vital signs tracking performance, we collected Xethru ground truth alongside CSI data for the environments 1, 2, 3 and 5.
We evaluate Serene's performance in terms of three key metrics: (1) breath tracking accuracy, (2) breathing signal outage (during which Serene cannot track breathing), and (3) naturally occurring motion false positives due to activity of other house occupants.
To determine long-term sleep quality metrics (\ie sleep efficiency based on sleep-awake classification discussed in \S \ref{sleepscoring}), we use data collected from the environments 1-4.
Based on these metrics, we present insights on sleep efficiency of different users and how it varies in successive nights.
Next, we first evaluate the breath tracking accuracy of our scheme.
 \presec
\subsection{Breath Tracking Accuracy}
\postsec
We evaluate the accuracy of Serene's breathing rate estimation in terms of \textit{BPM error}. 
We define BPM error as the average mean squared error (MSE) between per second BPM values reported by Serene and the corresponding ground truth BPM values reported by X4M200 over a specific time window.
We perform this evaluation on data collected from environments 1,2,3 and 5.
We evaluate both long-term (\ie whole night) and short-term (\ie specific short duration sleep windows during the night).

\presec
\subsubsection{Long-term Accuracy}
\textit{Serene achieves a median error of less than 1.19 BPM for real-world in-home full night sleep experiments.}
To evaluate Serene's long-term breath tracking accuracy, we compute the mean squared error (MSE) of instantaneous (per second) breathing rate estimate across an entire night of sleep.
The overall cumulative distribution function (CDF) of the MSE error in breath per minute (BPM) is depicted in figure~\ref{fig:eval2}. 
Inspecting the blue curve, we see that the median accuracy of Serene's breathing rate estimate is 1.19 BPM, while its 95th percentile confidence is under 1.9 BPM. 
We skip User 5's CDF from the graph as we were only able to collect 5 nights of data from that user. 
The average, minimum and maximum BPM errors observed for User 5 were 1.1, 0.811, and 1.14, respectively. 
Figure \ref{fig:evalall} shows how Serene tracks breathing rate throughout the night in for 4 different users, where we have plotted X4M200 ground truth side by side.
We can observe that BPM accuracies vary during the night as a user's sleep posture and distance from the sleep monitor can change during sleep.
However, we also observe that Serene is able to track the overall increasing and decreasing trend in a subject's breath during sleep fairly well when compared to Xethru's ground truth.

\begin{figure}[htbp]
    \centering
    \captionsetup{justification=centering}
    \subfigure[CDF of overall and per-user BPM error.]{
        \includegraphics[width=0.37\textwidth]{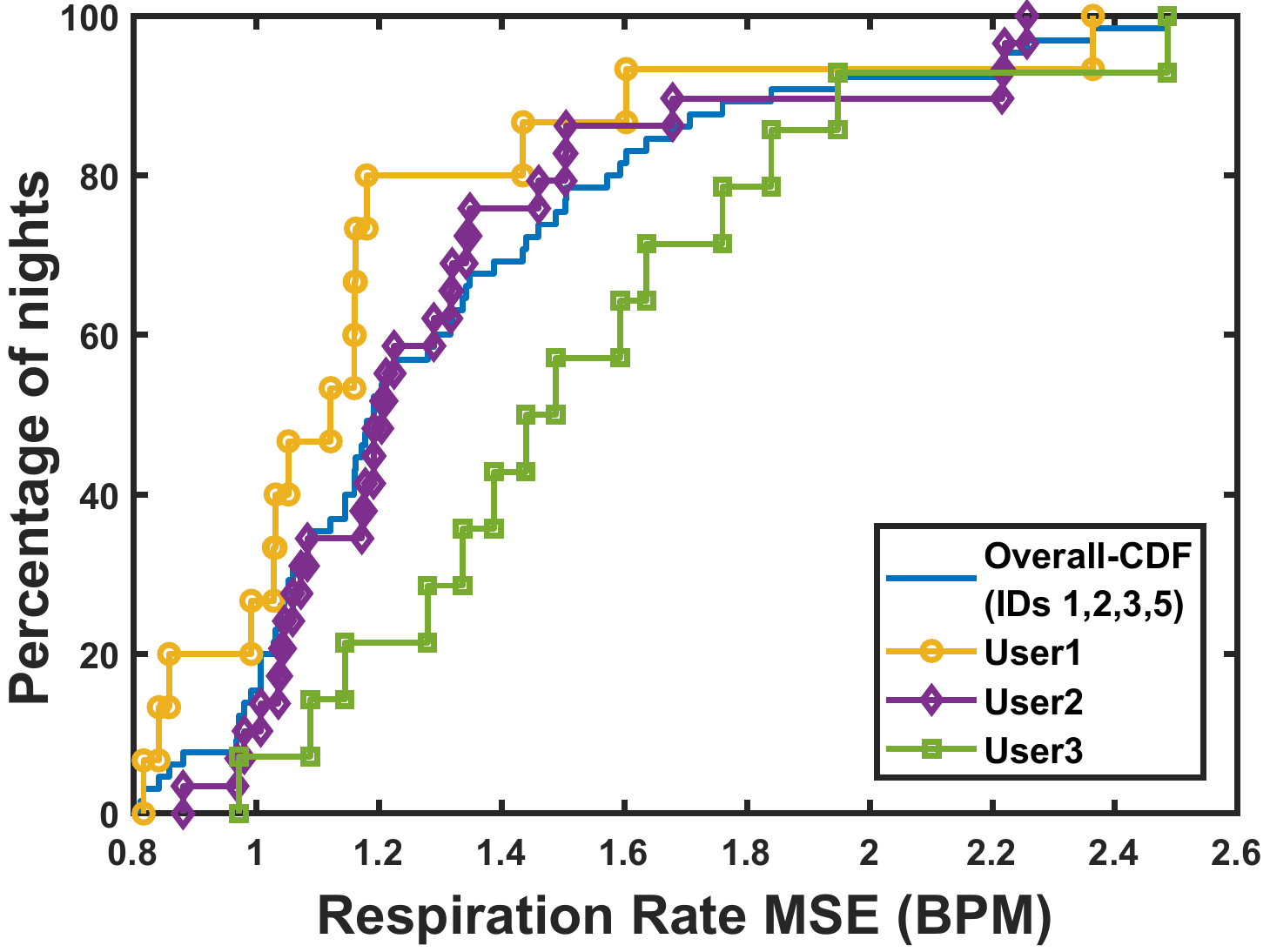}
        \label{fig:eval2}
    }
    \subfigure[Full night breath tracking.]{
        \includegraphics[width=0.5\textwidth]{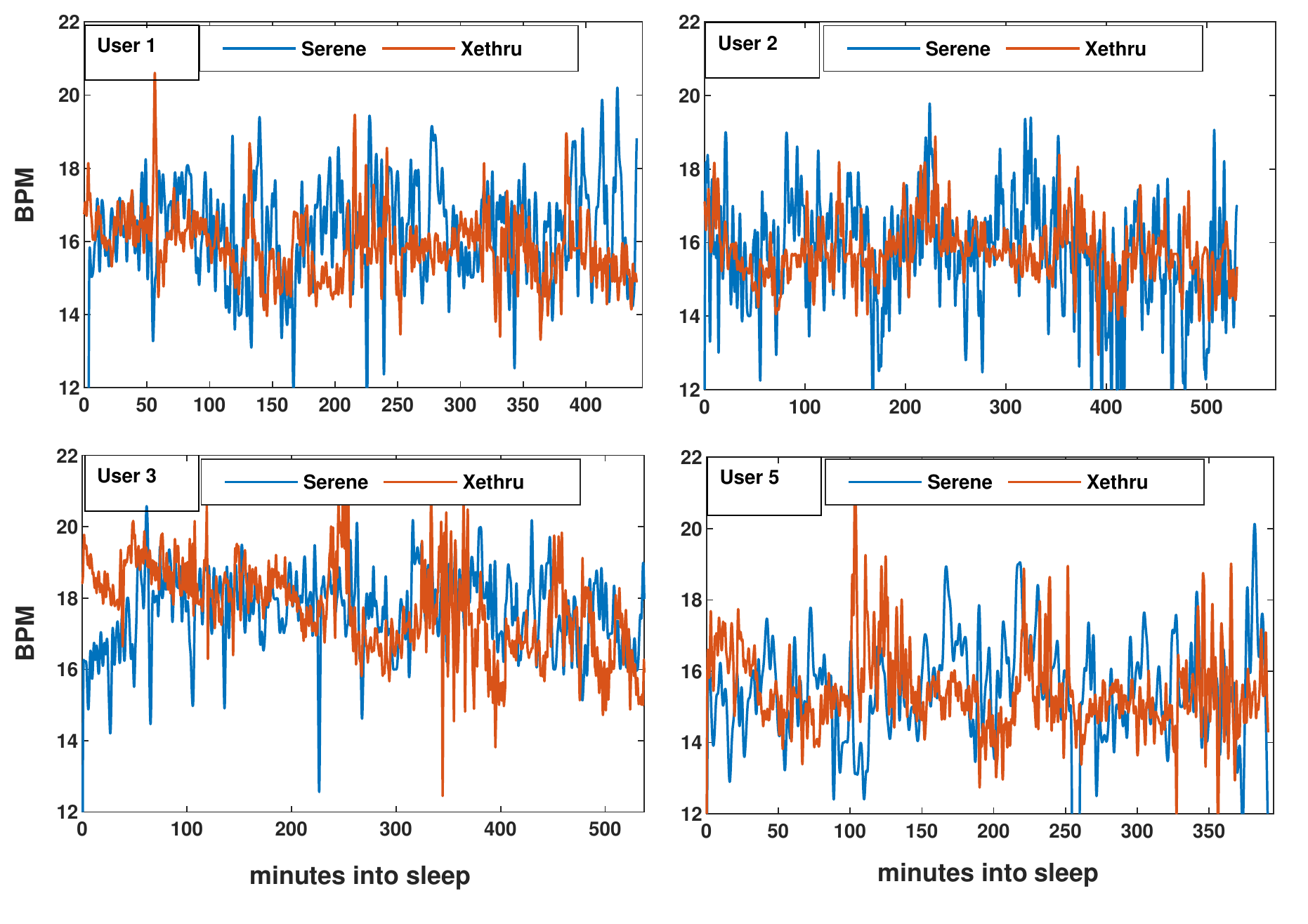}
        \label{fig:evalall}
    }
    \subfigure[Sleep posture experiments.]{
    \includegraphics[width=0.25\textwidth]{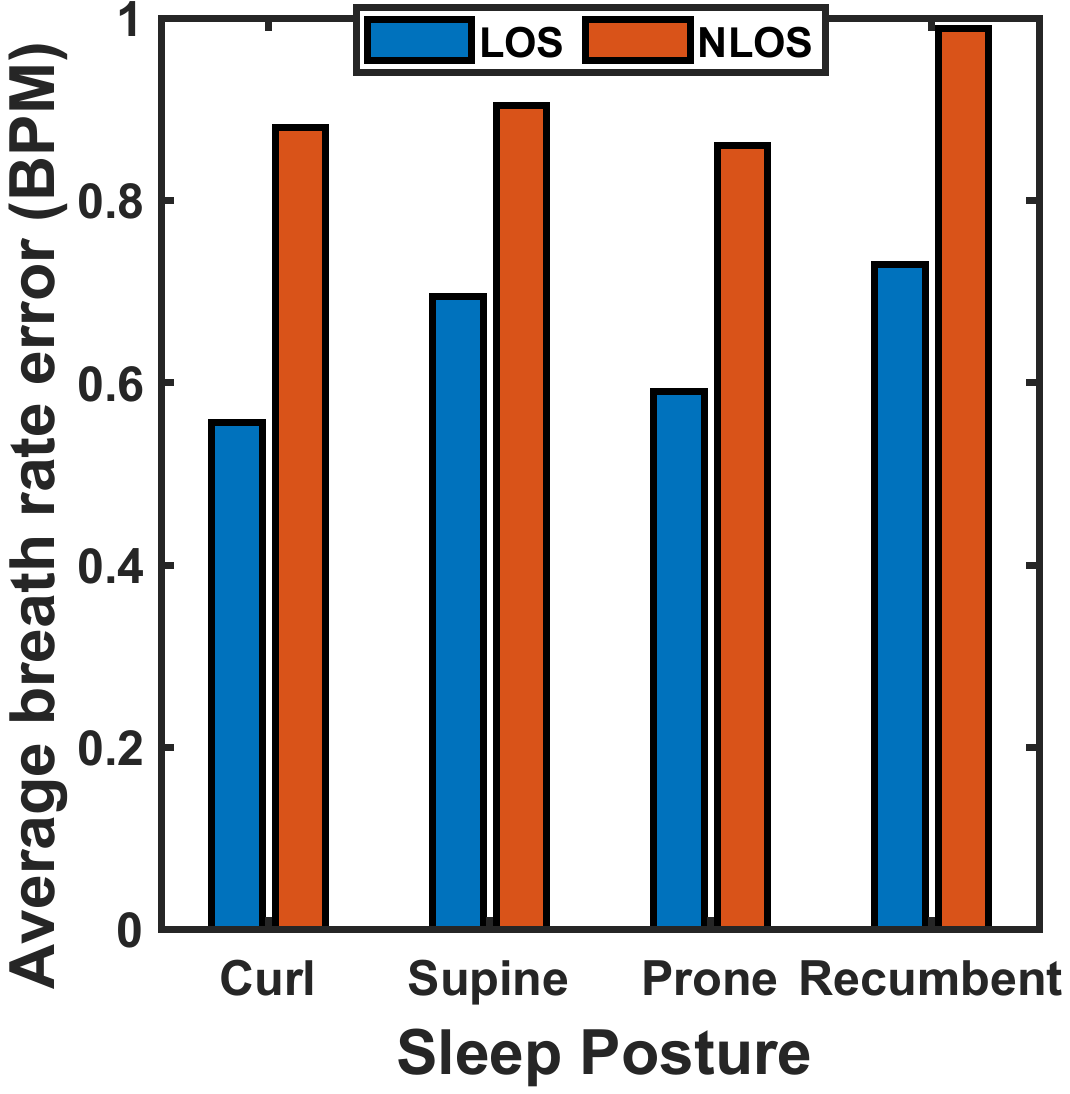}
    \label{fig:evalal3}
    }
    \vspace{-0.12in}
    \caption{CDF of overall and per-user breathing rate MSE compared to a Xethru X4M200 ground truth; Serene's full-night breath tracking performance; and average BPM errors for short duration sleep experiments in different sleep postures}
    \label{fig:evalbpmtracking}
    \vspace{-0.1in}
\end{figure}

Figure \ref{fig:eval2} also shows how BPM accuracy varies across subjects. 
For instance, the median accuracy was better than 1.12 BPM and 1.2 BPM for users 1 and 2, respectively. 
However, user 3's median accuracy was a bit higher (\ie 1.488), while the 95th percentile confidence was as large as 1.95 BPM.
These slight variations across different users and environments occur due to differences in physiques, respiratory system morphologies and environmental deployment conditions.
For example, environments 2 and 3 both correspond to NLOS scenarios, which leads to relatively lower BPM accuracies.
Although the level of robustness and accuracy Serene achieves may not be comparable to contact-based high accuracy breath monitors, yet it is comparable to other commercial contact-less sleep monitoring products.
Therefore, based on our results, we conclude that WiFi based sleep monitors can be robust and accurate enough for daily in-home use to gain insights into overall breathing trends during sleep. 
However, the accuracy may not be enough for medical grade sleep assessments.

\presec
\subsubsection{Short-term Accuracy} 
\textit{Serene can achieve an error of as low as 0.34 BPM during certain parts of a full-night sleep. However, the errors can be more than 5 BPM depending upon the time of night as a user's sleep posture and distance from the sleep monitor changes during sleep.}
In Fig. \ref{fig:evalall}, we notice that there are certain time windows during the night where Serene matches Xethru's performance very closely.
To know how many times such time windows occur during different nights in our dataset, we divide each night into small 15 minute time windows and compute the MSE of per second BPM estimate in those windows.
Figure \ref{fig:instantevals} show the CDF plots for 6 different full-night sleep experiments corresponding to users 1, 2 and 3.
From Fig. \ref{fig:instant1}, we observe that in the case of User 1, Serene experienced a breathing rate error of only 0.34 BPM in one 15 minute window during Night 6. 
Moreover, error in more than 50\% of the time windows remained below 0.84 BPM during Night 6.
Similarly, for other users, we observe that in multiple time windows during a full-night sleep, BPM error stays under 1 BPM.
This shows that Serene does fairly well when compared to controlled short-duration sleep experiments performed in recent CSI based sleep monitoring studies.
Also, figure \ref{fig:evalal3} shows average BPM errors for controlled 10 minute sleep experiments in different sleep postures.
We observe that Serene achieve an error of less than 1 BPM for most sleep postures even in NLOS scenarios.
The errors were as low 0.55 BPM in LOS scenarios.
However, we also observed errors approaching 5 BPM during certain time windows that can be attributed to changes in the user's sleep posture and distance from the sleep monitor during sleep.

\begin{figure}[htbp]
    \centering
    \captionsetup{justification=centering}
    \subfigure[User 1, 6 nights.]{
        \includegraphics[width=0.335\textwidth]{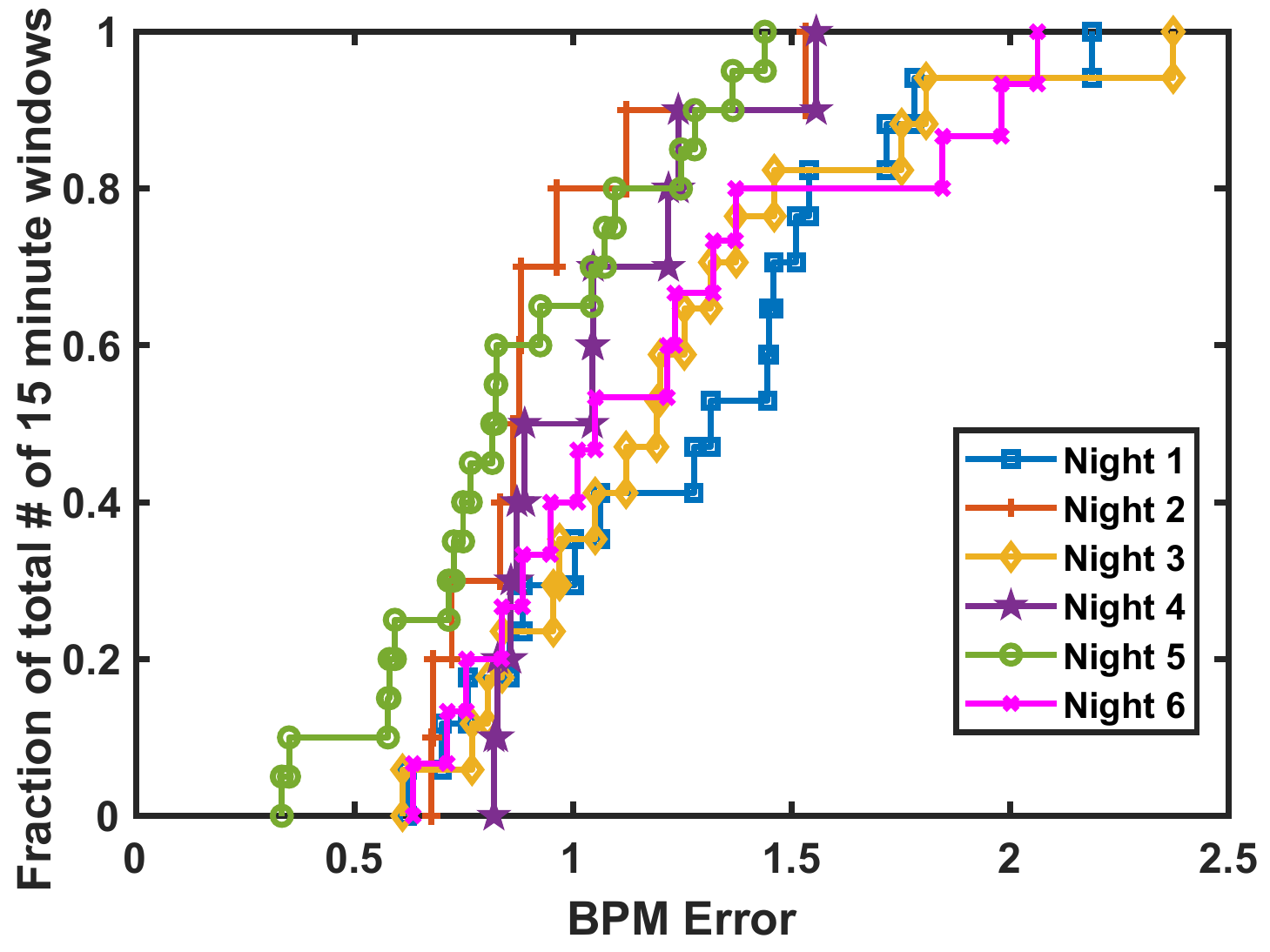}
        \label{fig:instant1}
    }
    \subfigure[User 2, 6 nights.]{
        \includegraphics[width=0.335\textwidth]{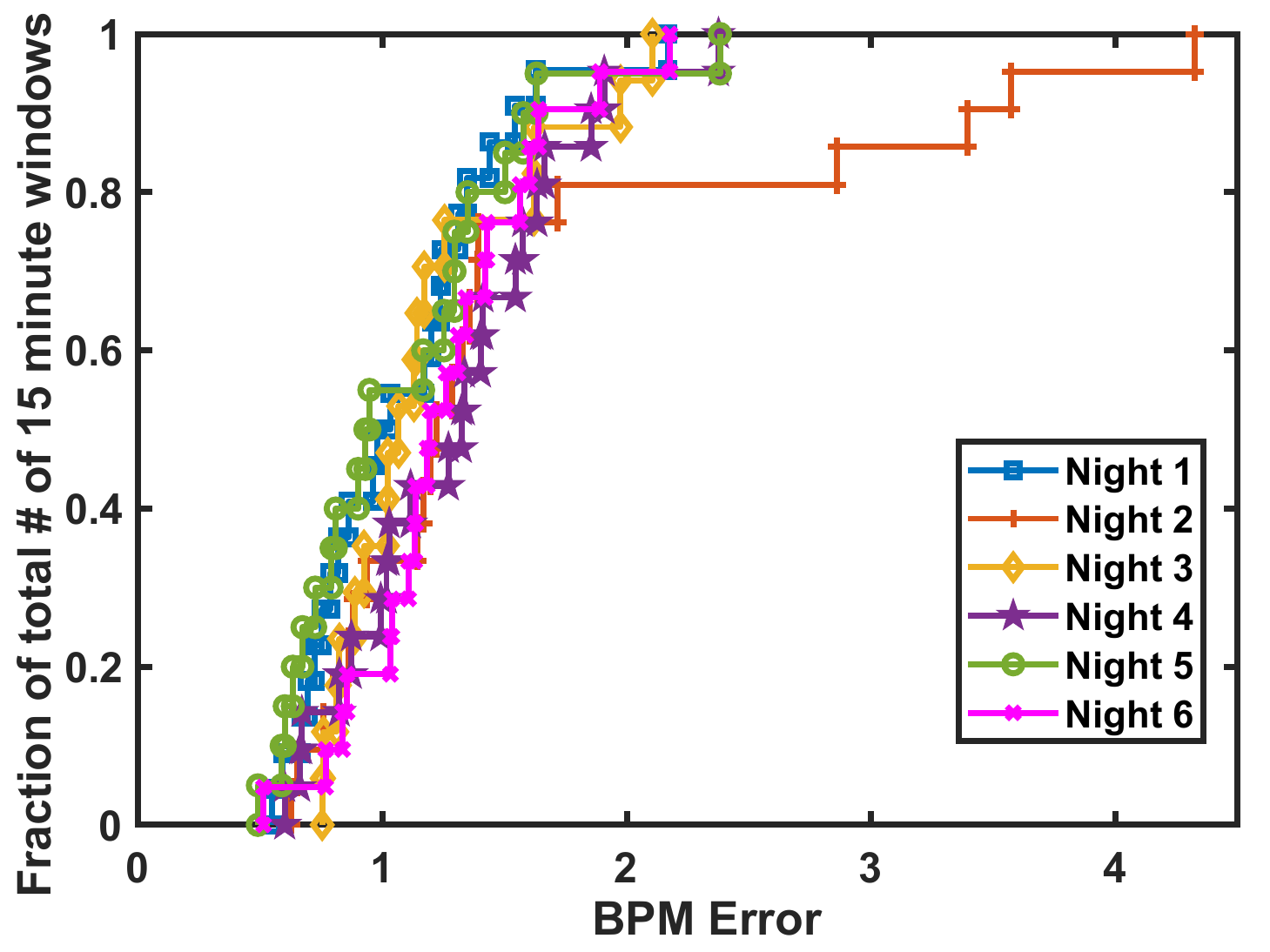}
        \label{fig:instant2}
    }
    \subfigure[User 3, 6 nights.]{
    \includegraphics[width=0.335\textwidth]{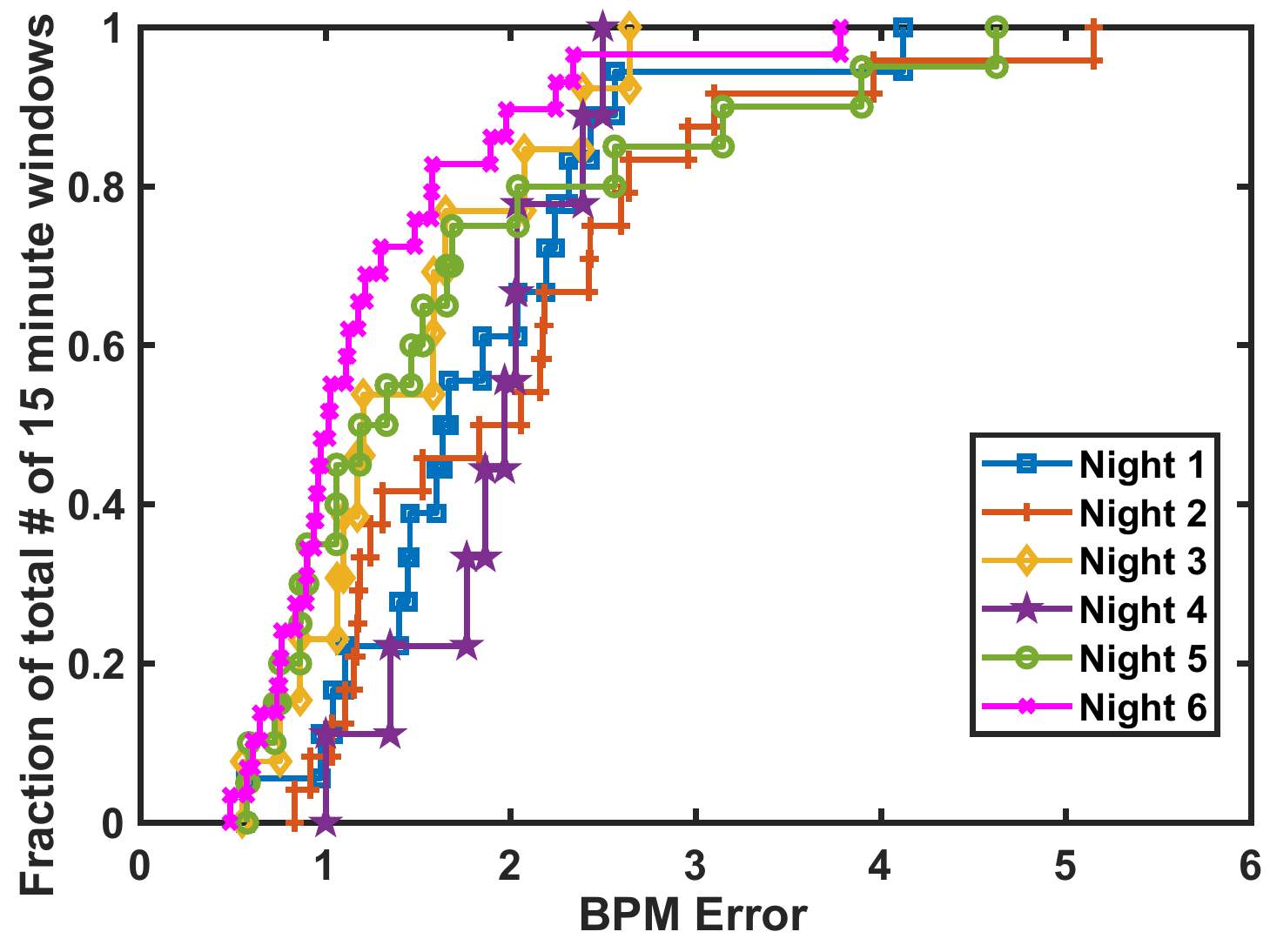}
    \label{fig:instant3}
}
    \vspace{-0.15in}
    \caption{CDFs of BPM errors calculated over 15 minute windows for 6 different nights (Users 1, 2 and 3).}
    \label{fig:instantevals}
    \vspace{-0.14in}
\end{figure}

 \presec
\subsection{Naturally Occurring Motion False Positives}
\postsec
\textit{Serene experienced a median of 20 false positive limb/body motion events, which can be attributed to activity of other house residents while the user is sleeping. The total duration of such events stayed below 37 minutes more than 95\% of the time.}
Radar and WiFi are both very sensitive to body motion.
We observed from our experiments that whenever a user moves in their bed, both Serene and X4M200 successfully detect the motion event.
However, we also observed scenarios where Serene detected body movements but the ground truth remained undisturbed (\ie contained breathing signal only).
We call such spurious movements detected by Serene as \textit{motion false positives} (MFPs), which we attribute to other movements present in the environment (e.g. when one of the occupants wakes up to get water, etc.).
To understand how significant such MFPs can be in real-world deployments of a WiFi based sleep monitoring system, we evaluate the following two key metrics on the real-life dataset we collected using Serene: (1) the \textit{number} and (2) \textit{duration} of MFPs per night's sleep.

\begin{figure}[htbp]
    \captionsetup{justification=centering}
    \subfigure[motion false positives]{
        \includegraphics[width=0.35\textwidth]{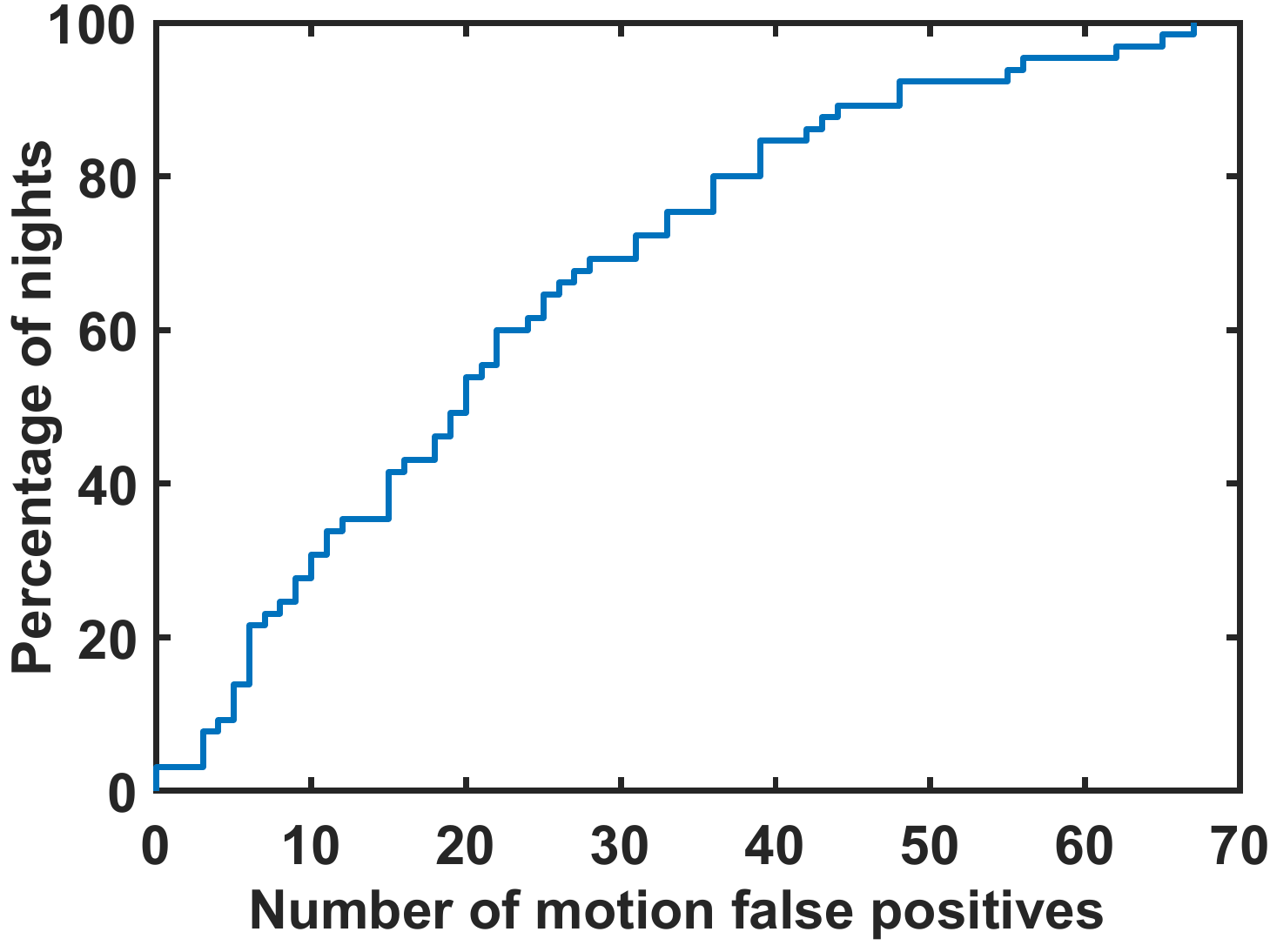}
        \label{fig:motionFalsePosNum}
    } %
    \subfigure[duration of motion false positives]{
        \includegraphics[width=0.35\textwidth]{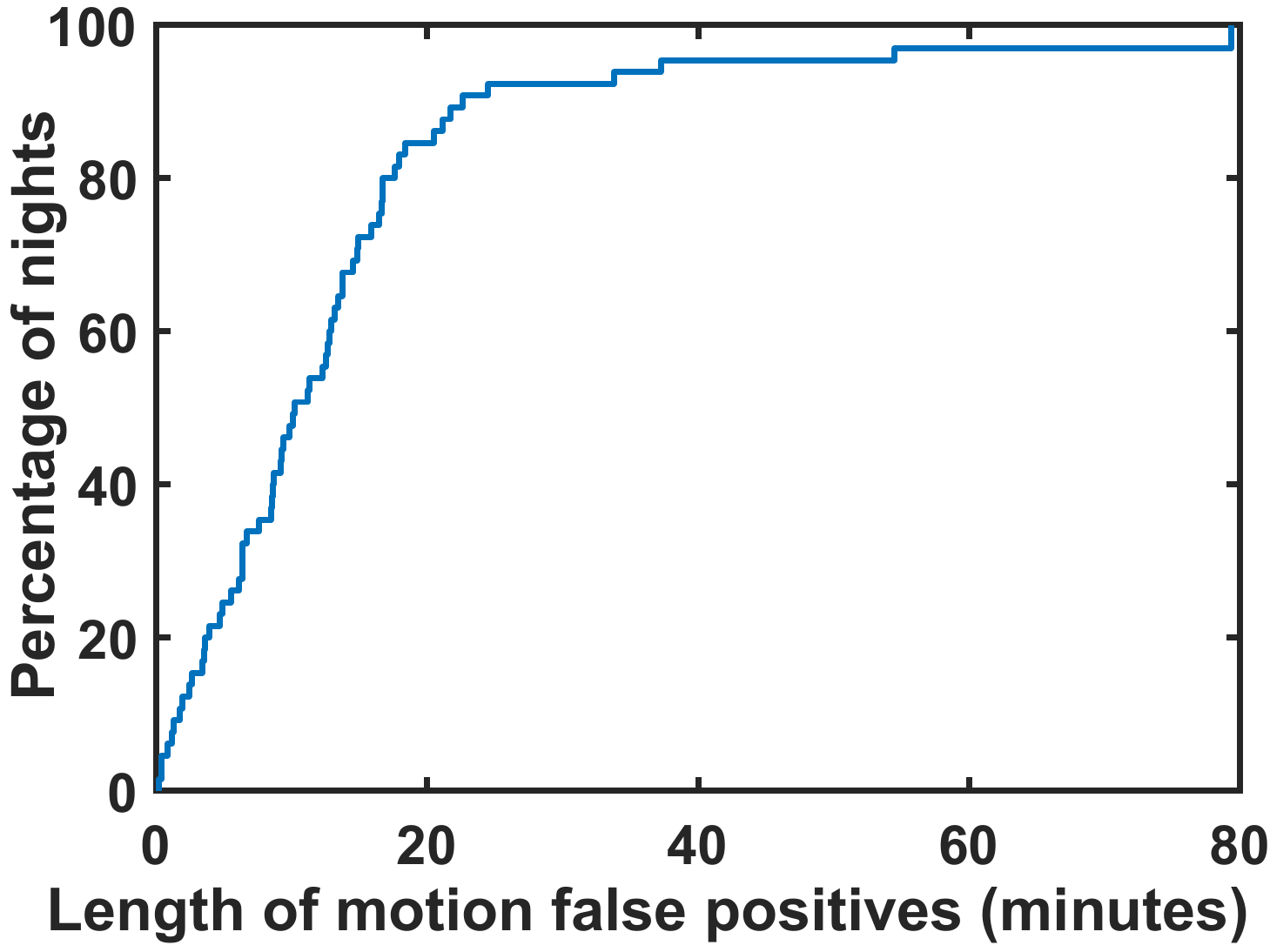}
        \label{fig:motionFalsePosDuration}
    }
    \vspace{-0.1in}
    \caption{CDF's of numbers and total duration of motion false positives during a night when compared with X4M200 ground truths. Motion false positive naturally occur due to activity of other housemates.}
    \label{fig:motion_stats}
    \vspace{-0.14in}
\end{figure}

Figure~\ref{fig:motion_stats} illustrates the CDFs of these two metrics evaluated over more than 65 nights in our database.
We observe that our system detected less than 56 MFPs occurrences for 95\% of tested nights~\ref{fig:motionFalsePosNum}.
Moreover, when we observe that when MFPs occur, their collective duration remains bounded under 37 minutes for 95\% of the time, as shown in figure~\ref{fig:motionFalsePosDuration}.
Although the total duration of such events during any night's stayed below 10 minutes on average and below 37 minutes 95\% of the time, yet we observed motion false positives of more than 60 minutes (1 hour) in total during one of the nights in our dataset.
Note that these MFPs do not signify any technical limitation of WiFi based sleep monitoring, as such motions occur naturally in home environments.
Our results show that the number and duration of naturally occurring interference in WiFi signals due to activity of other residents is usually low during night time.
Therefore, WiFi based sensing is suitable enough to be used for sleep monitoring during night time.

 \presub
\subsection{Breath Signal Outage} \label{sec:outage}
\postsub

\subsubsection{Detecting Outage in Breathing Signal} \textit{Serene experiences an average outage of less than 6.38 minutes, during which it cannot track any vital signs.} 
We define the \textit{outage} of our system as the event when variations due to breathing are not present in the CSI signal while the ground truth device (i.e. Xethru X4M200 in our case) is able to track breathing \footnote{To detect outage, we use the `absent' signal that Xethru X4M200 provides when it cannot detect any motion in the environment.}.
To detect outage, our system first determines the noise floor of the environment using the first PCA projection.
During real-time tracking, our system compares the average power of the signal, calculated over 7.5s windows of every 30s sleep epoch (i.e. $1/4^{th}$ of an epoch's duration), with the determined noise floor.
If the average power of the sleep epoch is not above the determined noise floor while X4M200 is still able to track breath, Serene signals outage.
To assess Serene's ability to continuously track vital signs (\ie breathing and other limb/body activity) in real life, we measure its per-night \textit{outage} performance statistics.
\begin{figure*}[htbp]
    \centering
    \captionsetup{justification=centering}
    \subfigure[Outage rate]{
        \includegraphics[width=0.31\textwidth]{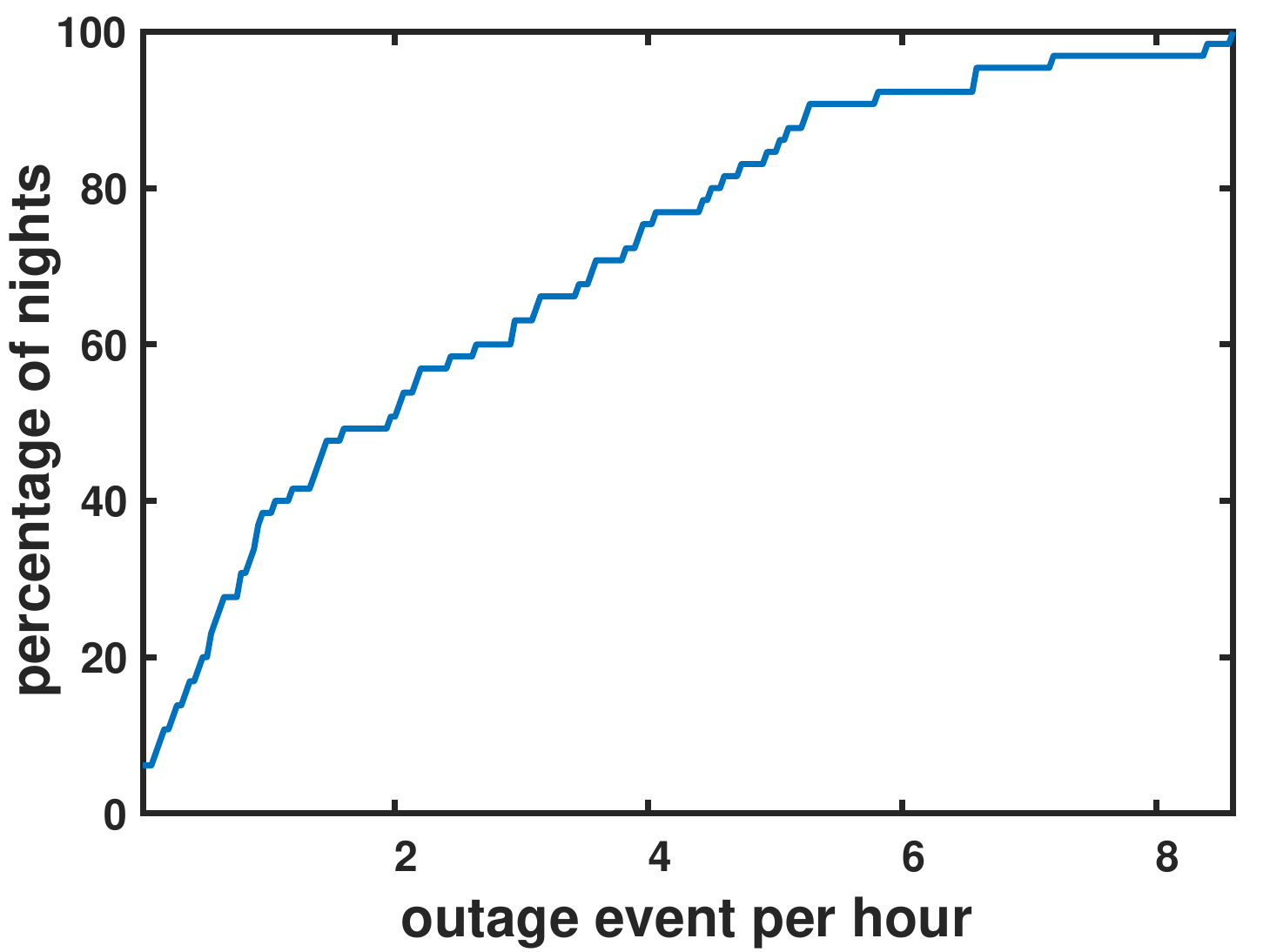}
        \label{fig:outage_rate}
    }
    \subfigure[small-scale outage duration]{
        \includegraphics[width=0.31\textwidth]{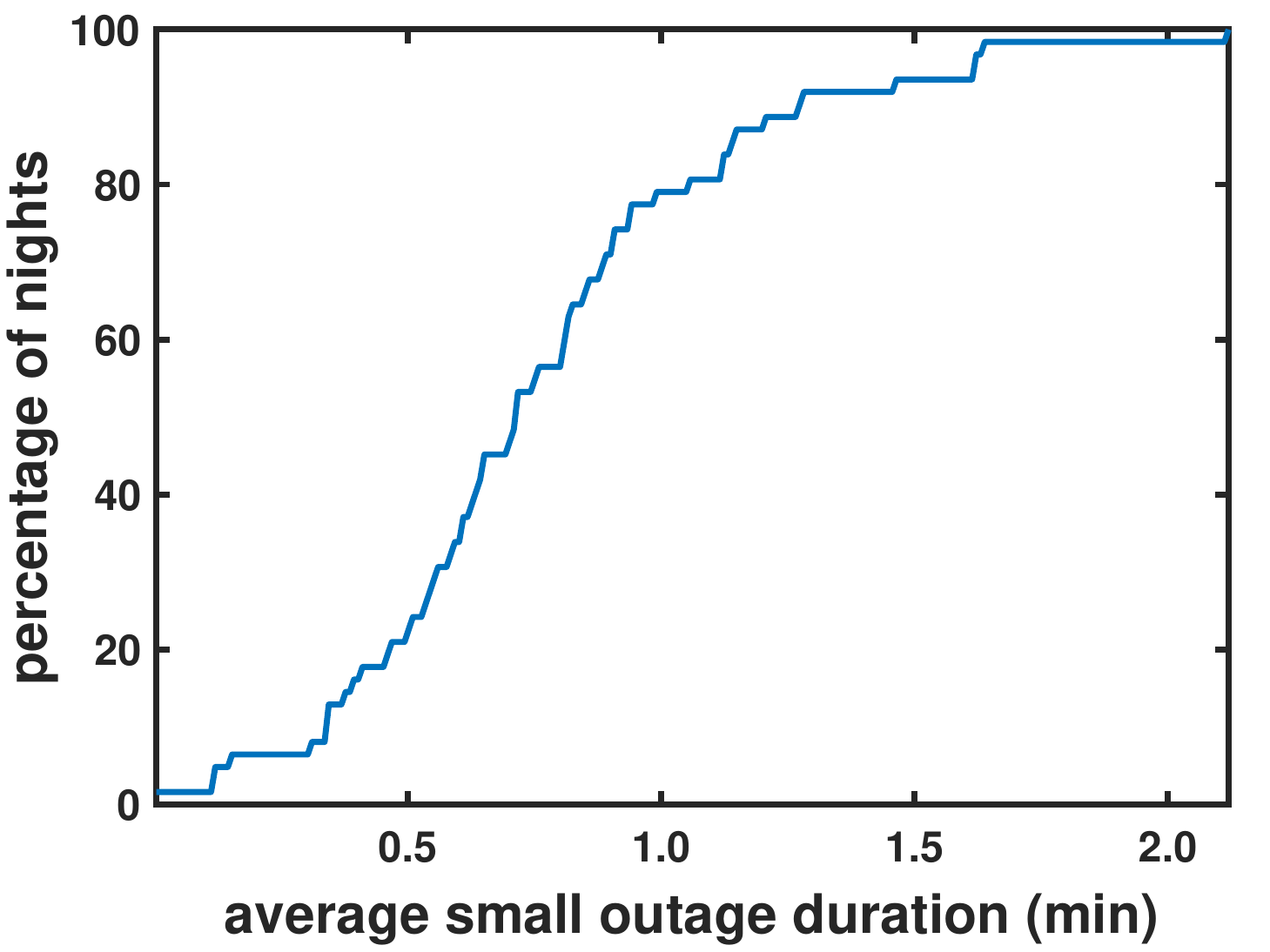}
        \label{fig:avrgOutageDurationSmall}
    }
    \subfigure[large-scale outage duration]{
        \includegraphics[width=0.31\textwidth]{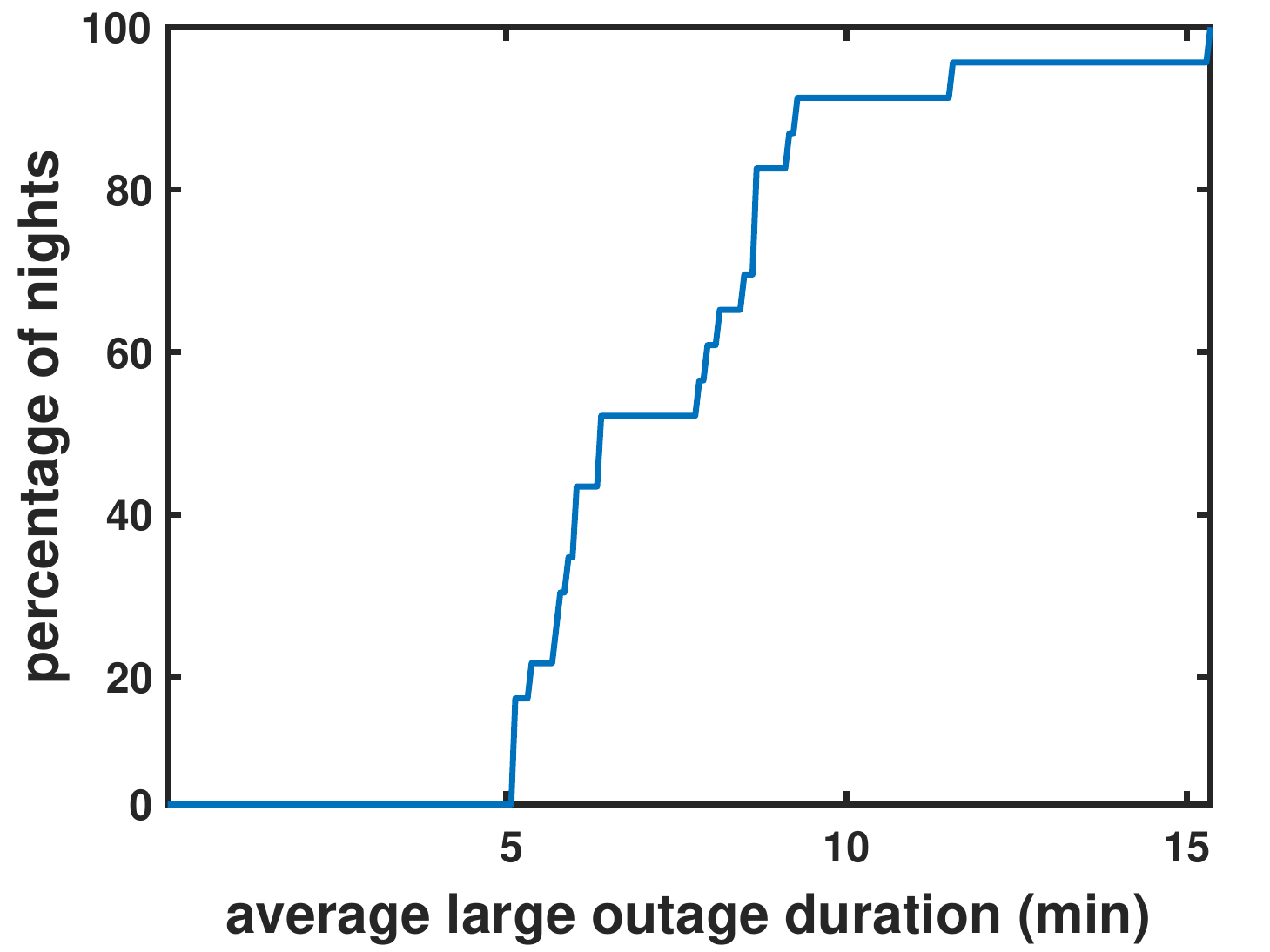}
        \label{fig:avrgOutageDurationLarge}
    }
    \vspace{-0.15in}
    \caption{Second-order statistics of breath estimation outage events. Outage rate and average outage duration mirror, respectively, their counterparts level crossing rate and average fade duration from wireless propagation literature.}
    \label{fig:outage_stats}
    \vspace{-0.13in}
\end{figure*}
To achieve this, we follow the treatment of signal outage in wireless propagation literature. %
Specifically, we calculate two second-order statistics: level crossing rate (LCR), and average fade duration (AFD)~\cite{Abdi00_ComparisonOfLcrAndAfd}.
LCR determines the rate at which outages occur during a full-night sleep, whereas AFD determines the duration of each outage.
We analyze the LCR and AFD using the first PCA projection's power with respect to the noise floor.
LCR and AFD have been extensively studied in body area network (BAN) literature owing to the complex and non-stationary way in which a human body interacts with the wireless channel~\cite{Smith15_ChannelModelingForWBAN}. %
Next, we present a summary of the aforementioned statistics derived from our entire dataset.
Figure~\ref{fig:outage_rate} shows LCR or outage rate calculated per hour across our sleep dataset. 
We can observe that on average, the breathing rate estimation of participants experienced 2 outage events per hour.
At 95 percentile confidence, outage amounted to less than 6.6 events per hour.
However, in the context of sleep monitoring, a further piece of detail must be considered to fully understand outage events during sleep, \ie the duration of such outages, which we characterize using AFD.
Serene can experience two types of outage events: \textit{small-scale} and/or \textit{large-scale}. 
Such outages arise due to users rolling over in bed to a different position or sleep posture while sleeping. 
This is because certain sleep postures can make it difficult for Serene to detect the breath signal due to weaker chest movements.
To understand how such small-scale and large-scale outage events are distributed naturally in real life, we introduce a design threshold to separate the two types of outage events. 
From the analysis of our dataset, we set such design threshold to 5 minutes, where we consider outage events longer than 5 minutes as a large-scale outages and vice versa. 
Figure~\ref{fig:avrgOutageDurationSmall} elaborates on the statistical behavior of small-scale outage. 
On average, small-scale outage events lasted for 0.7 minutes, while the 95th percentile confidence outage duration is under 1.62 minutes. 
CDF of the duration of large-scale outage is shown in figure~\ref{fig:avrgOutageDurationLarge}. 
Large-scale outage duration averaged around 6.38 minutes while its 95th percentile confidence is under 11.56 minutes, although durations in excess of 15 minutes can occur.
Based on this analysis, we can conclude that WiFi based sleep monitors experience more outages compared to radar based monitors such as Xethru X4M200.
However, we must mention that while collecting the dataset we suggested our subjects not to change the position of X4M200 so that their chest stays in X4M200's line-of-sight. 
In real life scenarios, users can make mistakes while positioning such radar based sensors before going to sleep which may cause outages similar to the ones experienced by Serene. 

 \presec
\section{Discussion} \label{sec:discussion}
\postsec

\subsection{Limitations of WiFi Signals based Vital Signs Monitoring during Sleep}

Our results show that WiFi based sleep monitoring can be significantly affected by changes in a user's sleep postures and activity of other house residents while the user is sleeping.
The breath rate error can vary between 0.34 BPM to more than 5 BPM depending upon the time of night as a user's sleep posture and distance from the sleep monitor can change during sleep.
Breath signal outages arise due to subjects rolling over in bed to a different position and/or sleep posture that makes it difficult to pick up the subjects' chest movements for a while.
False positive sleep motion events arise due to variations in CSI signals caused by other house residents of a sleeping user. 
However, average nightly duration of breath signal outages and motion false positives stayed under 6.38 and 10 minutes in our dataset, respectively.
Therefore, we conclude that WiFi based vital signs monitors perform fairly well compared to pulse-Doppler radar based solutions and can be robust and accurate enough for daily in-home use to gain insights into overall breathing and body movement trends during sleep. 
However, the accuracy may not be enough for medical grade sleep assessments.

\presec
\subsection{Sleep Scoring}\label{sleepscoring}
\postsec

\begin{figure*}[htbp]
    \centering
    \captionsetup{justification=centering}
    \includegraphics[width=1.01\textwidth]{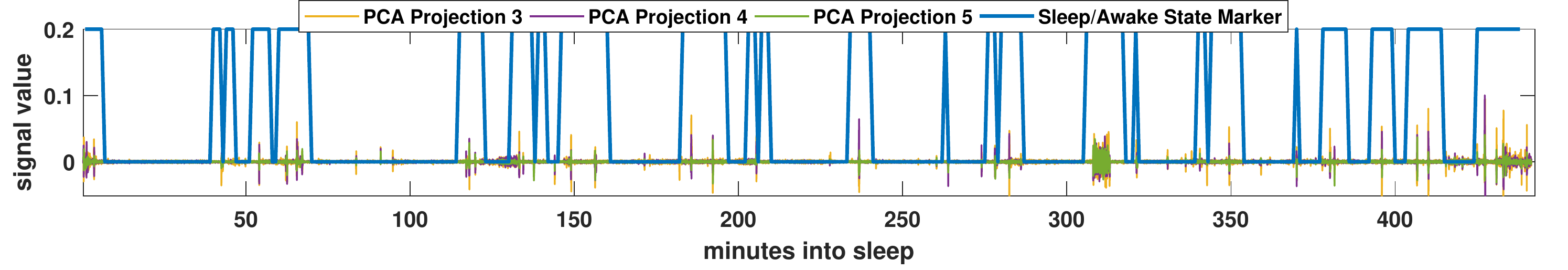}
    \vspace{-0.26in}
    \caption{Example showing Sleep/Awake classification for full night's sleep of a subject. Sleep efficiency was 62.1\%.}
    \label{fig:sleepawakescoring}
    \vspace{-0.13in}
\end{figure*}
\begin{figure*}[htbp]
    \centering
    \captionsetup{justification=centering}
    \includegraphics[width=\textwidth]{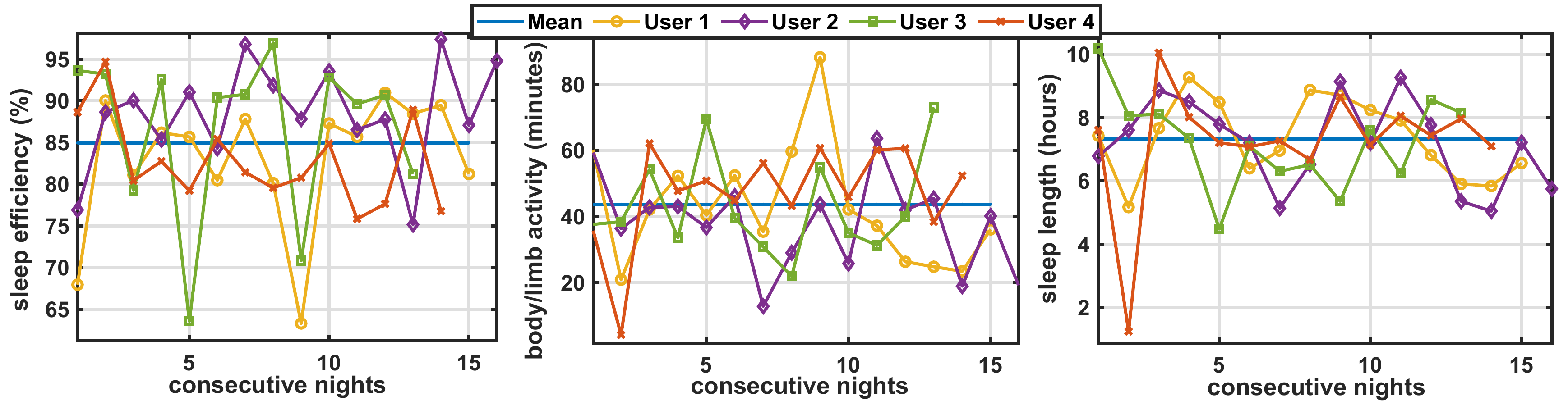}
    \vspace{-0.2in}
    \caption{Sleep efficiency and body motion corresponding to 4 users and throughout 13+ consecutive nights.}
    \label{fig:insight1}
    \vspace{-0.1in}
\end{figure*}
\begin{figure}[htbp]
    \centering
    \captionsetup{justification=centering}
    \subfigure[Overall and per-user CDF for motion duration.]{
        \includegraphics[width=0.36\textwidth]{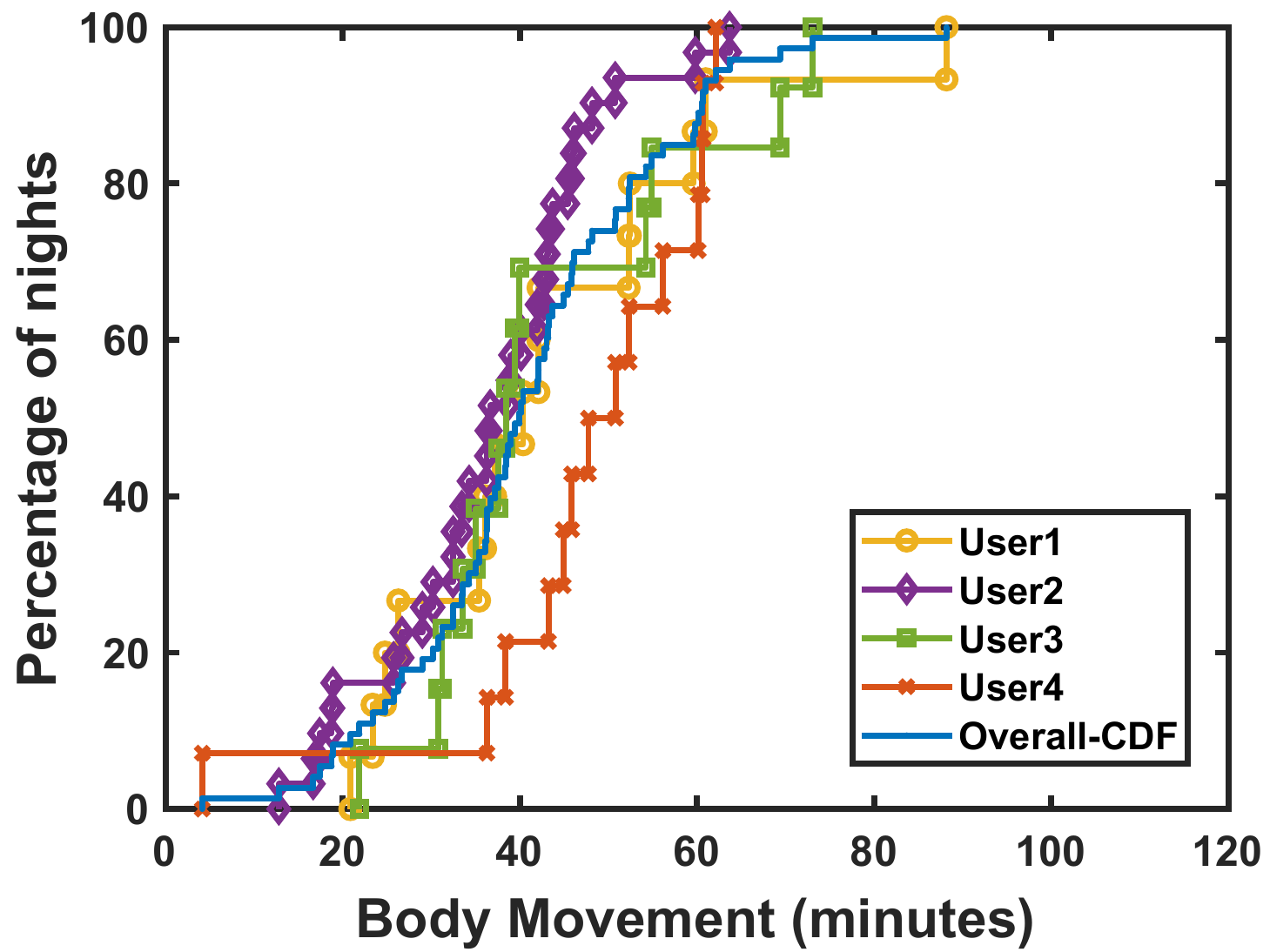}
        \label{fig:insight2}
    }
    \subfigure[Overall and per-user CCDF for sleep efficiency]{
        \includegraphics[width=0.36\textwidth]{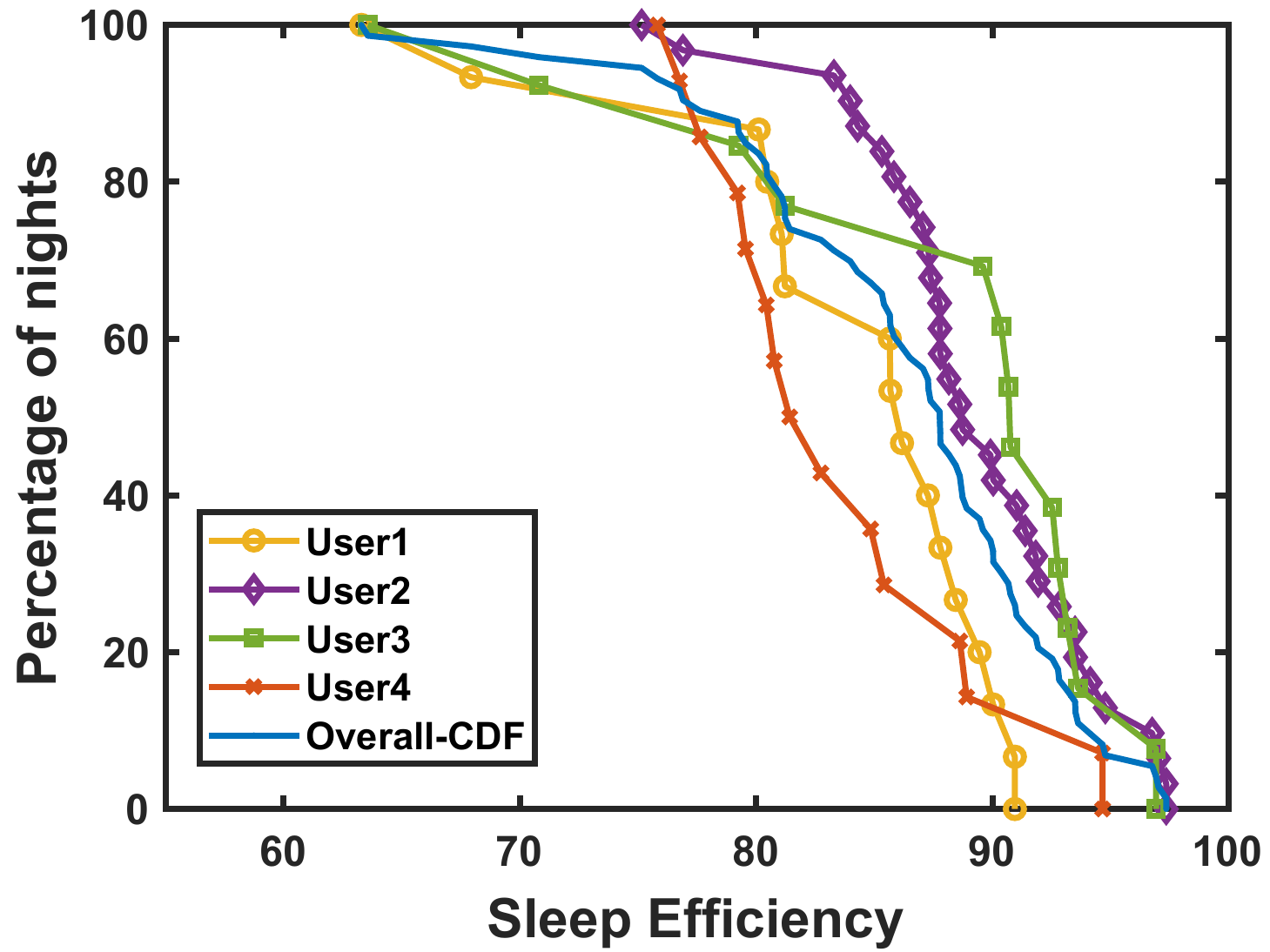}
        \label{fig:insight3}
    }
    \vspace{-0.15in}
    \caption{Overall and per-user CDF for motion duration and sleep efficiency.}
    \label{fig:insights}
    \vspace{-0.15in}
\end{figure}

To motivate the merits of WiFi based sleep monitoring, we present a few interesting insights on sleep quality gained from our data collection campaign.
To achieve this, we take an \textit{actigraphy} based approach towards sleep quality monitoring, where we classify the stage of each minute as \textit{sleep} or \textit{awake} period.
Our approach is inspired by the classic light-weight actigraphy based method proposed in \cite{webster1982activity}, which determines sleep-awake stage of a minute by taking into account body movement related information corresponding to the surrounding minutes.
The activity sleep-awake scores determined by their technique have been shown to agree with EEG based sleep monitoring 94.46\% of the time \cite{webster1982activity}.
In our implementation, we adopt the following model from their work, which takes 4 previous minutes and 2 following minutes into account to classify stage of the current minute:

\vspace{-0.1in}
\begin{equation}
\begin{split}
    s_m = \rho \times (w_{-4}a_{-4} + w_{-3}a_{-3} + w_{-2}a_{-2} + \\ w_{-1}a_{-1} + w_{0}a_{0} + w_{+1}a_{+1} + w_{+2}a_{+2})
\end{split}
\label{eq:sleepmodel}
\end{equation}
\vspace{-0.1in}

where $s_m$ is the average sleep-awake score for the current minute, $\rho$ is a scaling factor, $a_{-i}$, $a_{0}$, $a_{+i}$ are activity scores (normalized number of body movement events in each minute) for previous, current and following minutes, and $w_{-i}$, $w_{0}$, $w _{+i}$ represent weights for the previous minutes, current minute and following minutes. 
If $s_m \leq 1$, the current minute's stage is classified as \textit{sleep}, and if $s_m > 1$, the current minute's stage is classified as \textit{awake}.
In our implementation, we chose $\rho=0.125, w_{-4} = 0.15, w_{-3} = 0.15, w_{-2} = 0.15, w_{-1} = 0.08,  w_{0} = 0.21,  w_{1} = 0.12,  w_{2} = 0.13$, as suggested by the authors of \cite{webster1982activity} for best results in their real-world deployments. 

Next, we perform sleep assessments using the data corresponding to users 1 - 4.
Our results show how Serene can provide users with actionable feedback on a per-night basis towards the long-term tracking and management their sleeping habits based on the aforementioned sleep scoring algorithm.
Figure~\ref{fig:insight1} shows three different metrics of sleep determined for 3 users over a period of more than 13 consecutive days, namely \textit{sleep efficiency}, \textit{aggregate motion} (in minutes) during sleep and \textit{sleep length}.
Sleep efficiency for each night of sleep was calculated using on our actigraphy based sleep scoring approach, which is defined as
the ratio of actual time spent in sleep stages to total time spent in bed (i.e. $T_{sleep}/(T_{sleep} + T_{awake})$).
Figure \ref{fig:sleepawakescoring} shows Sleep/Awake classification performance for full night's sleep of a subject, where sleep efficiency was determined to be 62.1\%.
As users manually started and ended each night's data collection using our software, the sleep lengths were easily determined according to those end points.
We observe interesting insights for these long term sleep metrics.
For instance, we can see that User 1 experienced a noticeably restless 9th night which resulted in poor sleep efficiency.
User 4 only slept for 1.25 hours, but as he was awake for only 4.156 minutes during that time, his sleep efficiency reaches ~95\%.

In terms of aggregate body motion statistics over nights and across subjects, we measured a median of ~40 minutes with the 95th percentile being under 80 minutes as illustrated in the blue CDF in figure~\ref{fig:insight2}. 
On an individual basis, and considering user 2 and user 4 for instance, their median body movements were ~36 minutes and ~47 minutes, respectively. 
This insight is corroborated when inspecting the complementary CDF's depicted in figure~\ref{fig:insight3}. 
Specifically, while both users 2 and 3 have a comparable maximum sleep efficiency of ~96\%, User 3's sleep efficiency was lower than 80\% on 3 different nights. 
Moreover, User 2 has a worst efficiency of ~75\%, whereas User 3 has worst efficiency of ~63\%. 
For the aggregate dataset, the median user population sleep efficiency was around 87\%.
The average sleep duration among these 3 users during this consecutive testing period was 7.32 hours.
Note that the recommended sleep for ages 18-64 years is 7-9 hours \cite{long2015analysis}.

 \presec
\section{Conclusions}
\postsec
In this paper, we evaluate the performance of WiFi based vital signs monitoring in the wild.
We make two major contributions.
First, we characterize the relationship between WiFi signal components (i.e. multipath and signal subspace) and human vital signs (i.e. respiration and body motions).
Grounded in this characterization, we propose two methods: 1) a respiration tracking technique that models the peak dynamics observed in the time-varying signal subspaces and 2) a body-motion tracking technique built with a multi-dimensional clustering of evolving signal subspaces.
Second, we extensively evaluate our proposed methods through real-world full-night sleep experiments conducted in 5 different apartments, where we collected more than $>$~550 hours (80 nights) of data from 5 users.
Our results demonstrate that the proposed techniques were able to track respiration rate with an average error of \textless1.19 breaths per minute (BPM).
However, the breath rate error varied between 0.34 BPM to more than 5 BPM depending upon the time of night as a user's sleep posture and distance from the sleep monitor can change during sleep.
Co-located activity of other house residents also affects WiFi based vital signs monitoring.
For example, our system experienced 20 false positive motion events on average every night, due to activity from co-located house residents while a user is sleeping. The duration of time during which respiration monitoring halted (i.e. estimation outage) was under 10 minutes on average per night.
We conclude that WiFi based vital signs monitors can be robust and accurate enough for daily in-home use to gain insights into overall breathing trends during sleep. 
However, the accuracy may not be enough for medical grade sleep assessments.

}

{ %\small
    \bibliographystyle{unsrt}
    \bibliography{SleepTech,WiFiTech,SleepMed,OtherTech}

\begin{thebibliography}{10}

\bibitem{sadeh1995role}
Avi Sadeh, Peter~J Hauri, Daniel~F Kripke, and Peretz Lavie.
\newblock The role of actigraphy in the evaluation of sleep disorders.
\newblock {\em Sleep}, 1995.

\bibitem{sadeh2002role}
Avi Sadeh and Christine Acebo.
\newblock The role of actigraphy in sleep medicine.
\newblock {\em Sleep medicine reviews}, 2002.

\bibitem{ancoli2003role}
Sonia Ancoli-Israel, Roger Cole, Cathy Alessi, Mark Chambers, William
  Moorcroft, and Charles~P Pollak.
\newblock The role of actigraphy in the study of sleep and circadian rhythms.
\newblock {\em Sleep}, 2003.

\bibitem{long2014sleep}
Xi~Long, Pedro Fonseca, J{\'e}r{\^o}me Foussier, Reinder Haakma, and Ronald~M
  Aarts.
\newblock Sleep and wake classification with actigraphy and respiratory effort
  using dynamic warping.
\newblock {\em IEEE journal of biomedical and health informatics}, 2014.

\bibitem{fitbit}
Fitbit.
\newblock Fitbit.
\newblock https://www.fitbit.com/, July 2018.

\bibitem{miband}
Xiaomi.
\newblock Xiaomi mi band 3.
\newblock https://www.mi.com/en/miband/, July 2018.

\bibitem{ouraring}
O.
\newblock Oura ring.
\newblock https://ouraring.com/, July 2018.

\bibitem{zeo}
Gibson~Research Corporation.
\newblock Zeo sleep manager pro.
\newblock
  https://uk.pcmag.com/zeo-sleep-manager-pro/5064/review/zeo-sleep-manager-pro,
  December 2012.

\bibitem{choe2010opportunities}
Eun~Kyoung Choe, Julie~A Kientz, Sajanee Halko, Amanda Fonville, Dawn
  Sakaguchi, and Nathaniel~F Watson.
\newblock Opportunities for computing to support healthy sleep behavior.
\newblock In {\em ACM CHI}, 2010.

\bibitem{hao2013isleep}
Tian Hao, Guoliang Xing, and Gang Zhou.
\newblock isleep: unobtrusive sleep quality monitoring using smartphones.
\newblock In {\em \Proc of ACM Sensys}, 2013.

\bibitem{pevernagie2010acoustics}
Dirk Pevernagie, Ronald~M Aarts, and Micheline De~Meyer.
\newblock The acoustics of snoring.
\newblock {\em Sleep medicine reviews}, 2010.

\bibitem{de2009detection}
GR~De~Bruijne, PCW Sommen, and RM~Aarts.
\newblock Detection of epileptic seizures through audio classification.
\newblock In {\em 4th European conference of the \Inte Federation for Medical
  and Biological Engineering}. Springer, 2009.

\bibitem{heinrich2015video}
Adrienne Heinrich, Frank van Heesch, Bhargava Puvvula, and Mukul Rocque.
\newblock Video based actigraphy and breathing monitoring from the bedside
  table of shared beds.
\newblock {\em Journal of Ambient Intelligence and Humanized Computing}, 2015.

\bibitem{poh2011advancements}
Ming-Zher Poh, Daniel~J McDuff, and Rosalind~W Picard.
\newblock Advancements in noncontact, multiparameter physiological measurements
  using a webcam.
\newblock {\em IEEE Transactions on Biomedical Engineering}, 2011.

\bibitem{migliorini2010automatic}
Matteo Migliorini, Anna~M Bianchi, Domenico Nistic{\`o}, Juha Kortelainen,
  Edgar Arce-Santana, Sergio Cerutti, and Martin~O Mendez.
\newblock Automatic sleep staging based on ballistocardiographic signals
  recorded through bed sensors.
\newblock In {\em IEEE EMBC}, 2010.

\bibitem{paalasmaa2012unobtrusive}
Joonas Paalasmaa, Mikko Waris, Hannu Toivonen, Lasse Lepp{\"a}korpi, and Markku
  Partinen.
\newblock Unobtrusive online monitoring of sleep at home.
\newblock In {\em IEEE EMBC}, 2012.

\bibitem{kortelainen2010sleep}
Juha~M Kortelainen, Martin~O Mendez, Anna~Maria Bianchi, Matteo Matteucci, and
  Sergio Cerutti.
\newblock Sleep staging based on signals acquired through bed sensor.
\newblock {\em IEEE Transactions on Information Technology in Biomedicine},
  2010.

\bibitem{chen2005unconstrained}
Wenxi Chen, Xin Zhu, Tetsu Nemoto, Yumi Kanemitsu, Keiichiro Kitamura, and
  Ken-ichi Yamakoshi.
\newblock Unconstrained detection of respiration rhythm and pulse rate with one
  under-pillow sensor during sleep.
\newblock {\em Medical and Biological Engineering and Computing}, 2005.

\bibitem{choi2009slow}
Byung~Hun Choi, Gih~Sung Chung, Jin-Seong Lee, Do-Un Jeong, and Kwang~Suk Park.
\newblock Slow-wave sleep estimation on a load-cell-installed bed: a
  non-constrained method.
\newblock {\em Physiological measurement}, 2009.

\bibitem{withingssleeppad}
Withings.
\newblock Withings sleep tracking mat.
\newblock https://www.withings.com/us/en/sleep, 2018.

\bibitem{rahman2015dopplesleep}
Tauhidur Rahman, Alexander~T Adams, Ruth~Vinisha Ravichandran, Mi~Zhang,
  Shwetak~N Patel, Julie~A Kientz, and Tanzeem Choudhury.
\newblock Dopplesleep: A contactless unobtrusive sleep sensing system using
  short-range doppler radar.
\newblock In {\em \Proc of ACM Ubicomp}, 2015.

\bibitem{liu2014wi}
Xuefeng Liu, Jiannong Cao, Shaojie Tang, and Jiaqi Wen.
\newblock Wi-sleep: Contactless sleep monitoring via wifi signals.
\newblock In {\em IEEE RTSS}, 2014.

\bibitem{liu2015tracking}
Jian Liu, Yan Wang, Yingying Chen, Jie Yang, Xu~Chen, and Jerry Cheng.
\newblock Tracking vital signs during sleep leveraging off-the-shelf wifi.
\newblock In {\em \Proc of ACM MobiHoc}, 2015.

\bibitem{yang2016monitoring}
Zhicheng Yang, Parth~H Pathak, Yunze Zeng, Xixi Liran, and Prasant Mohapatra.
\newblock Monitoring vital signs using millimeter wave.
\newblock In {\em \Proc of ACM MobiHoc}, 2016.

\bibitem{adib2015smart}
Fadel Adib, Hongzi Mao, Zachary Kabelac, Dina Katabi, and Robert~C Miller.
\newblock Smart homes that monitor breathing and heart rate.
\newblock In {\em \Proc of ACM CHI}, 2015.

\bibitem{yue2018extracting}
Shichao Yue, Hao He, Hao Wang, Hariharan Rahul, and Dina Katabi.
\newblock Extracting multi-person respiration from entangled rf signals.
\newblock {\em \Proc of ACM IMWUT}, 2018.

\bibitem{occhiuzzi2010rfid}
Cecilia Occhiuzzi and Gaetano Marrocco.
\newblock The rfid technology for neurosciences: feasibility of limbs'
  monitoring in sleep diseases.
\newblock {\em IEEE Transactions on Information Technology in Biomedicine},
  2010.

\bibitem{occhiuzzi2014night}
Cecilia Occhiuzzi, Carmen Vallese, Sara Amendola, Sabina Manzari, and Gaetano
  Marrocco.
\newblock Night-care: A passive rfid system for remote monitoring and control
  of overnight living environment.
\newblock {\em Elsevier Procedia Computer Science}, 2014.

\bibitem{hillyard2018experience}
Peter Hillyard, Anh Luong, Alemayehu~Solomon Abrar, Neal Patwari, Krishna
  Sundar, Robert Farney, Jason Burch, Christina Porucznik, and Sarah~Hatch
  Pollard.
\newblock Experience: Cross-technology radio respiratory monitoring performance
  study.
\newblock In {\em ACM MOBICOM}, 2018.

\bibitem{wang2016human}
Hao Wang, Daqing Zhang, Junyi Ma, Yasha Wang, Yuxiang Wang, Dan Wu, Tao Gu, and
  Bing Xie.
\newblock Human respiration detection with commodity wifi devices: do user
  location and body orientation matter?
\newblock In {\em \Proc of ACM Ubicomp}, 2016.

\bibitem{zhang2018fresnel}
Fusang Zhang, Daqing Zhang, Jie Xiong, Hao Wang, Kai Niu, Beihong Jin, and
  Yuxiang Wang.
\newblock From fresnel diffraction model to fine-grained human respiration
  sensing with commodity wi-fi devices.
\newblock {\em ACM IMWUT}, 2018.

\bibitem{long2015analysis}
Xi~Long.
\newblock On the analysis and classification of sleep stages from
  cardiorespiratory activity.
\newblock {\em SLEEP-WAKE}, 2015.

\bibitem{dafna2015sleep}
Eliran Dafna, Ariel Tarasiuk, and Yaniv Zigel.
\newblock Sleep-wake evaluation from whole-night non-contact audio recordings
  of breathing sounds.
\newblock {\em PloS one}, 2015.

\bibitem{nguyen2017libs}
Anh Nguyen, Raghda Alqurashi, Zohreh Raghebi, Farnoush Banaei-Kashani, Ann~C
  Halbower, and Tam Vu.
\newblock Libs: A lightweight and inexpensive in-ear sensing system for
  automatic whole-night sleep stage monitoring.
\newblock {\em GetMobile: Mobile Computing and Communications}, 2017.

\bibitem{sun2017sleepmonitor}
Xiao Sun, Li~Qiu, Yibo Wu, Yeming Tang, and Guohong Cao.
\newblock Sleepmonitor: Monitoring respiratory rate and body position during
  sleep using smartwatch.
\newblock {\em \Proc of ACM IMWUT}, 2017.

\bibitem{kay2012lullaby}
Matthew Kay, Eun~Kyoung Choe, Jesse Shepherd, Benjamin Greenstein, Nathaniel
  Watson, Sunny Consolvo, and Julie~A Kientz.
\newblock Lullaby: a capture \& access system for understanding the sleep
  environment.
\newblock In {\em \Proc of ACM Ubicomp}, 2012.

\bibitem{gu2014intelligent}
Weixi Gu, Zheng Yang, Longfei Shangguan, Wei Sun, Kun Jin, and Yunhao Liu.
\newblock Intelligent sleep stage mining service with smartphones.
\newblock In {\em \Proc of ACM Ubicomp}, 2014.

\bibitem{min2014toss}
Jun-Ki Min, Afsaneh Doryab, Jason Wiese, Shahriyar Amini, John Zimmerman, and
  Jason~I Hong.
\newblock Toss'n'turn: smartphone as sleep and sleep quality detector.
\newblock In {\em \Proc of ACM CHI}, 2014.

\bibitem{zhao2017learning}
Mingmin Zhao, Shichao Yue, Dina Katabi, Tommi~S Jaakkola, and Matt~T Bianchi.
\newblock Learning sleep stages from radio signals: a conditional adversarial
  architecture.
\newblock In {\em IEEE ICML}, 2017.

\bibitem{halperin2011tool}
Daniel Halperin, Wenjun Hu, Anmol Sheth, and David Wetherall.
\newblock Tool release: gathering 802.11 n traces with channel state
  information.
\newblock {\em ACM SIGCOMM Computer Communication Review}, 2011.

\bibitem{Alloulah18_SubspaceTrackingWifi}
M.~{Alloulah}, A.~{Isopoussu}, C.~{Min}, and F.~{Kawsar}.
\newblock {On Tracking the Physicality of Wi-Fi: A Subspace Approach}.
\newblock {\em IEEE Access}, 7:19965--19978, 2019.

\bibitem{Mosht2011}
M.~Moshtaghi and et. al.
\newblock Incremental elliptical boundary estimation for anomaly detection in
  wireless sensor networks.
\newblock In {\em IEEE ICDM}, 2011.

\bibitem{xethru200}
Xethru.
\newblock Respiration sensor x4m200.
\newblock https://www.xethru.com/x4m200-respiration-sensor.html, 2018.

\bibitem{hummingboard}
SolidRun.
\newblock Hummingboard.
\newblock https://www.solid-run.com/nxp-family/hummingboard/, 2018.

\bibitem{xethrureliability}
Xethru.
\newblock Xethru vs. polysomnography (psg) comparative study.
\newblock
  https://www.xethru.com/community/resources/categories/white-papers.6/.

\bibitem{Abdi00_ComparisonOfLcrAndAfd}
Ali Abdi, Kyle Wills, H~Allen Barger, M-S Alouini, and Mostafa Kaveh.
\newblock Comparison of the level crossing rate and average fade duration of
  rayleigh, rice and nakagami fading models with mobile channel data.
\newblock In {\em Vehicular Technology \Conf, 2000. IEEE-VTS Fall VTC 2000.
  52nd}, volume~4, pages 1850--1857. IEEE, 2000.

\bibitem{Smith15_ChannelModelingForWBAN}
David~B Smith and Leif~W Hanlen.
\newblock Channel modeling for wireless body area networks.
\newblock In {\em Ultra-Low-Power Short-Range Radios}, pages 25--55. Springer,
  2015.

\bibitem{webster1982activity}
John~B Webster, Daniel~F Kripke, Sam Messin, Daniel~J Mullaney, and Grant
  Wyborney.
\newblock An activity-based sleep monitor system for ambulatory use.
\newblock {\em Sleep}, 1982.

\end{thebibliography}
}

\balance

\end{document}